# Discovery of a Phenomenological Dynamical Model for Predicting the El Niño-Southern Oscillation

**Authors:** Shivsai Ajit Dixit[1*], B. N. Goswami[2]

**Affiliations:**

[1]Indian Institute of Tropical Meteorology, Pune.

[2]Cotton University, Guwahati.

*Correspondence to: sadixit@tropmet.res.in

**Abstract**:

The skill of the statistical as well as physics-based coupled climate models in predicting the El Niño-Southern Oscillation (ENSO) is limited by their inability to represent the observed ENSO nonlinearity. A promising alternative, namely a deterministic nonlinear dynamical model derived from an observed ENSO timeseries, however, has remained elusive. Here we discover such a phenomenological nonlinear dynamical model that embodies known physical processes responsible for the self-sustained quasi-oscillatory character of the ENSO and its observed spectrum of variability. High predictive potential of the model is demonstrated and the intrinsic nonlinearity of the ENSO is shown to be critical for overcoming the Spring Predictability Barrier to a large extent. The unique methodology presented here has the potential for constructing similar models for other geophysical systems.

**One Sentence Summary:** A deterministic nonlinear model, with high predictive potential and little computational cost, is extricated from an observed ENSO time series.

**Main Text:**

The El Niño–Southern Oscillation (ENSO) phenomenon being the largest signal of year-to-year climate fluctuation on our planet arising from tropical ocean and atmosphere interactions (*1,2*) has impacts that are felt worldwide in both natural systems and human affairs (*3*). Since the first failed forecast of ENSO (*4,5*) followed by the first successful forecast a decade later (*6*), the ENSO forecasts have seen tremendous progress (*7*) - thanks to availability of better observations and modeling tools (both physical and statistical). That the ENSO forecasts still remains a major challenge is illustrated by the grand failure of all statistical and dynamical models to forecast the 2014 event ahead of its development (*8*). Although ENSO events have an affinity to be locked with the seasonal cycle and has a dominant periodicity of recurrence, the predictability is limited due to the inherent aperiodicity arising from the nature of the ENSO with no cycle repeating itself exactly. While the long range predictability of ENSO comes from its quasi-cyclic component, the limit on the predictability of ENSO arises from the faster growth of errors due to the inherent nonlinearity (*9*). The asymmetry of ENSO (strong El Niño Events and weaker La Niña events) is a manifestation of the intrinsic nonlinearity (*10,11*). The challenge of ENSO prediction is further raised a notch higher by the existence of the diversity of the ENSO namely, the existence of the so-called CP (Central Pacific) El Niño and EP (Eastern Pacific) El Niño (*12-14*) which again appears to be a result of interaction of the ENSO nonlinearity with the background mean state (*15,16*).

Correct representation of the observed nonlinearity and its interaction with mean background state in a model, therefore, is critical for improving the skill of ENSO prediction by the model. Notwithstanding tremendous progress made by coupled dynamical models in simulating the ENSO (*17*), all models still suffer from significant biases in simulating the mean ocean and atmospheric climate over the tropical Pacific. The current limited skill of the dynamical models (*18,7*) is related to the inability of the models to represent the nonlinearity and aperiodicity with fidelity as a result of these biases of the models in simulating the mean. This is also the reason for the statistical models' skill being comparable to that of the dynamical models (*7*). Linear and nonlinear inverse models of ENSO prediction with stochastic forcing (*19*) may reproduce some statistics of the observed ENSO time series but do not incorporate the process nonlinearity explicitly. With the improvement of the coupled dynamical models and reduction of their biases in simulating the mean climate, the skill of the dynamical models is expected to improve steadily. However, this process is going to be slow and the statistical models could still provide ENSO forecasts comparable to or better than those by the dynamical models if these models could incorporate the observed nonlinearity and aperiodicity with fidelity. Due to very little operational cost of such forecasts, they could be an ideal independent addition to the dynamical forecasts for multi-model ensemble forecasts of ENSO.

The observed nonlinearity and aperiodicity is contained in an ENSO index, for example as defined by the so-called MEI.ext index (*20*); see Fig.1. How do we model the underlying dynamics of such a time series accurately preserving the nonlinearity and aperiodicity? Our overarching objective in this study is to derive the essential nonlinearity of the ENSO from observations and derive a nonlinear dynamical model describing an ENSO time series with fidelity. Towards this end, an ENSO index time series is identified. Defining ENSO variability with periods longer than one year as the ENSO 'slow manifold', we use nonlinear dynamical concepts to identify the nonlinearity in the ENSO and discover a time-invariant, deterministic, nonlinear phenomenological model that describes the evolution of the ENSO slow manifold to an excellent approximation. We demonstrate that the nonlinear model is equivalent to a nonlinear differential equation that describes the underlying dynamics of ENSO. This finding is a major new insight about ENSO dynamics and to our knowledge, the first-of-its-kind, that has been derived entirely from observations. We further demonstrate that the nonlinear model could be used to make useful hindcast predictions of ENSO with lead time up to 10.5 months using past data from the ENSO time series alone. The skill of the nonlinear predictive model has a weak seasonality with only a mild Spring Predictability Barrier (SPB).

**MEI.ext index timeseries and the ENSO slow manifold**

Predictability of the ENSO has its origins in the slow-varying quasi-cyclic component (slow manifold) arising from coupled ocean-atmosphere interactions. From amongst various indices devised for monitoring the ENSO phenomenon, here we use an index called MEI.ext (denoted by $\widetilde{y}$, Fig. 1A) that incorporates the intrinsic coupled character of ENSO (*20*) and spans a long period of 135 years from January 1871 ($t=0$) to December 2005 ($t=1619$) with effectively one datapoint every month (see Materials and Methods in SM).

Next, by filtering $\widetilde{y}$ to remove climate noise (periodicities shorter than one year), we obtain the *slow manifold* (denoted by $y$, Fig. 1A) which captures all important features of the ENSO variability and retains most of the energy (about 95%) of $\tilde{y}$ (Fig. 1B). Henceforth in this paper we shall work with $y$ unless specified otherwise.

An integral timescale $T$, intrinsic to the slow manifold, can be derived based on the first zero-crossing ($\tau = 14$ months, Fig. 1C) of the autocorrelation coefficient plotted against delay time $\tau$

$$T = \int_{\tau=0}^{\tau=14} R d\tau. \tag{1}$$

$T$ represents the shortest average time over which $y$ is correlated with itself (21); here $T = 7$ months (Fig. 1C). All temporal quantities may then be nondimensionalized as $\hat{t} = t/T$, $\hat{\tau} = \tau/T$ etc.

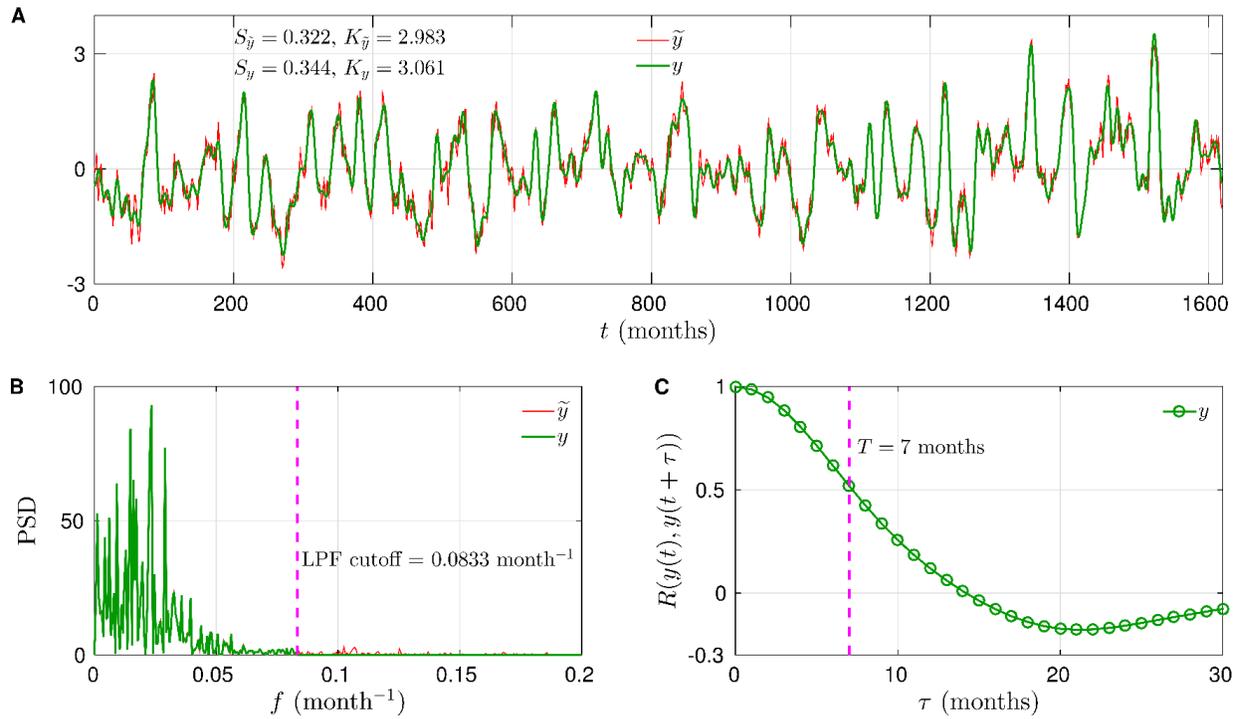

**Fig. 1. Timeseries of the ENSO index MEI.ext and the ENSO slow manifold. A,** Timeseries of the ENSO index MEI.ext (*20*) denoted as $\tilde{y}$. MEI.ext extends from January 1871 ($t=0$) to December 2005 ($t=1619$) i.e. 135 years with effectively one datapoint every month i.e. 1620 datapoints. ENSO slow manifold is denoted by $y$. The slow manifold $y$ is obtained by spectral filtering of $\tilde{y}$ using FFT. Corresponding skewness factor (*S*) and kurtosis (*K*) values are also shown. **B,** Power spectral density (PSD) of $y$ and $\tilde{y}$. Energy of the signal is area under the PSD

curve. Location of the low-pass spectral filter is shown by a vertical dashed line. **C,** Autocorrelation coefficient $R$ of $y$ plotted against the time delay $\tau$. Integral timescale $T$ (Eq. 1) is shown by the vertical dashed line.

**A nonlinear predictive model for the ENSO slow manifold**

One method of reconstructing the $m$-dimensional phase space of a dynamical system is the so-called time-delay embedding (*22-25*) wherein $m$ copies of the timeseries of an observable are created with each copy delayed from the earlier one by time delay $\tau$. The corresponding datapoints from all copies form $m$ coordinates of the state point in the phase space. While some earlier studies have shown that ENSO dynamics may be explained by a lower-order nonlinear model (*24,25*), other studies have argued for (*26*) as well as against (*27*) chaos in ENSO dynamics; all these studies typically use $m = 6$. In this work, we use $m = 5$ and *a posteriori* find it entirely sufficient to describe the evolution of the ENSO slow manifold. We take the view that ENSO slow manifold dynamics contains important nonlinearities (*24,25*) but is not in the chaotic regime (*27*) possibly due to the specific values of the parameters involved (*28,29*). Given this, the standard procedures - devised for chaotic attractors - for deciding upon the correct embedding dimension $m$ (*30*) and time delay $\tau$ (*23*), need not be strictly followed here. For instance, it is required to maintain $\tau \ll T$ for arriving at the nonlinear ODE underlying the ENSO slow manifold (see Materials and Methods in SM) and make useful predictions, as opposed to $\tau > T$ for capturing the chaotic attractor, if present.

Figure 2(A) shows the building block for an ENSO slow manifold state point in $m = 5$ phase space. As the dynamics unfolds in time, it is possible that the state point moves along the phase-space trajectory in such a way that a certain functional relationship, possibly nonlinear and presumably deterministic, between the coordinates of the state point holds for all times i.e. a time-invariant description might be possible. It turns out that such functional relationship indeed holds to an excellent approximation in the present case (see Materials and Methods in SM) and is of the form

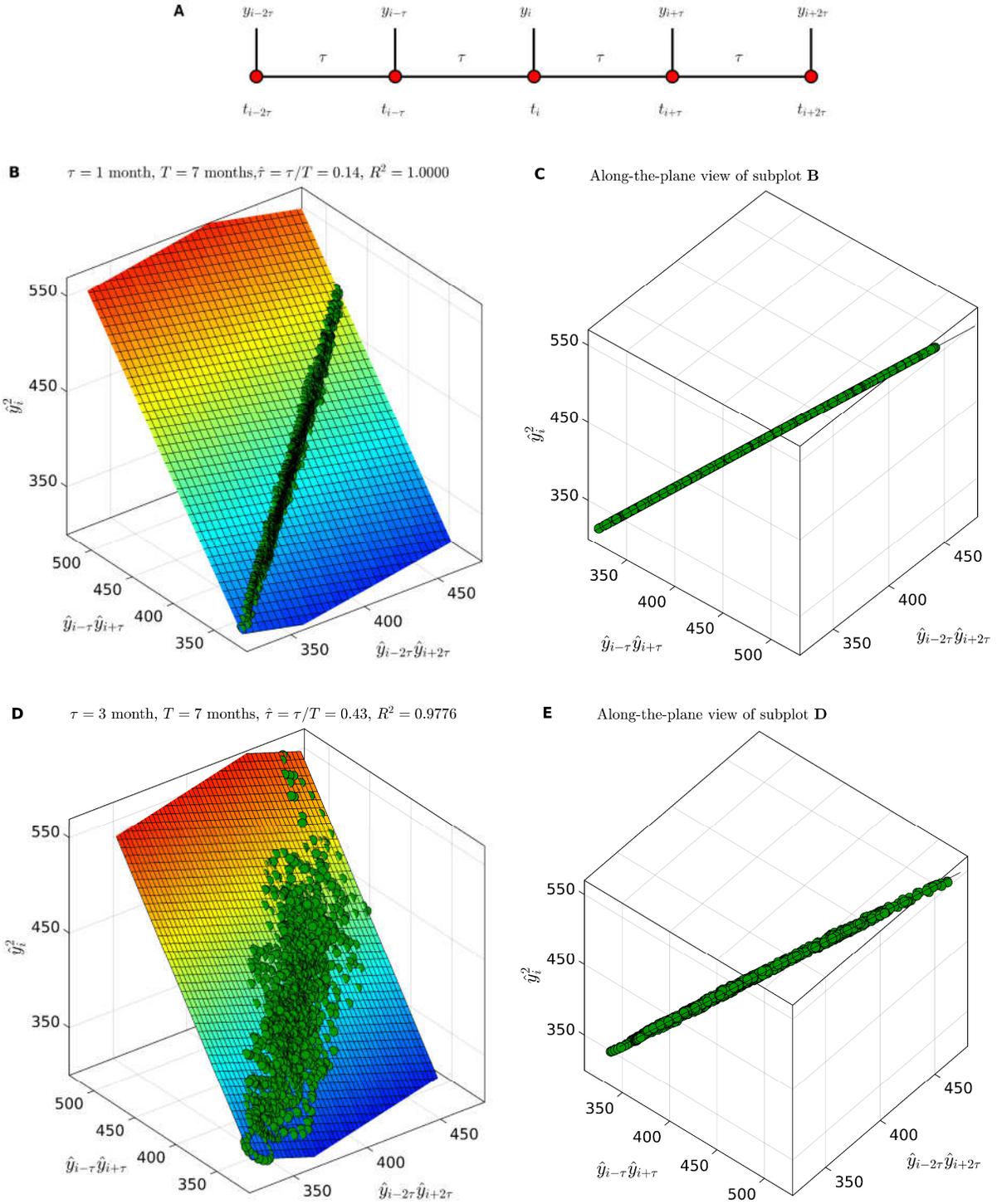

**Fig. 2. Time-invariant, deterministic, nonlinear model of the ENSO slow manifold** $y$. **A,** Building block of five points, spaced $\tau$ apart on the time axis, resulting from embedding $y$, with time delay $\tau$, in the five-dimensional phase space ($m = 5$). Temporal evolution of the slow manifold is related to the evolution of the building block. **B-E,** A reduced three-dimensional description of the five-dimensional phase space of $y$. Note that $y$ has been DC-shifted by 20 (i.e. $\hat{y} = y + 20$). Each point in the reduced space has coordinates $\hat{y}_{i-2\tau}\hat{y}_{i+2\tau}$, $\hat{y}_{i-\tau}\hat{y}_{i+\tau}$ and $\hat{y}_i^2$ and represents the state of the ENSO slow manifold. Trajectory of the slow manifold is shown for dimensionless time delays of **B,** $\hat{\tau} = 0.14$ and **D,** $\hat{\tau} = 0.43$. The trajectory, in each case, remains remarkably confined to a flat plane (fitted by the least-squares method) given by Eq. (2) as evident from the high $R^2$ values of the fit; coefficients $A$, $B$ and $C$ of the fitted plane (Eq. 2) are functions of $\hat{\tau}$. **C** and **E** show along-the-plane views of **B** and **D** respectively.

$$\hat{y}_i^2 = A + B\hat{y}_{i-2\tau}\hat{y}_{i+2\tau} + C\hat{y}_{i-\tau}\hat{y}_{i+\tau}, \qquad (2)$$

where $A$, $B$ and $C$ are coefficients that depend on the delay time $\tau$. Note that $\hat{y}$ in Eq. (2) is $\hat{y} = y + S$ where $S$ is a simple DC shift added to $y$ (see Materials and Methods in SM); Supplementary Material (Fig. S1) shows that the DC shift is crucial for obtaining improved hindcast skill of the nonlinear model. Equation (2) is akin to the generalized Volterra series used to describe time-invariant, nonlinear relationship between input and output of a dynamical system (*31*). Figures 2(B-E) show the trajectory of the ENSO slow manifold using $\hat{y}$ data ($S = 20$), embedded in the five-dimensional phase space, plotted in the reduced three-dimensional space with coordinates $\hat{y}_{i-2\tau}\hat{y}_{i+2\tau}$, $\hat{y}_{i-\tau}\hat{y}_{i+\tau}$ and $\hat{y}_i^2$ (see Eq. 2). Two delay times, $\tau = 1$ month and $\tau = 3$ months, are shown. It is evident that as the system evolves in time, the dynamics remarkably unfolds on a flat plane in the reduced space i.e. the system trajectory is tightly confined to the flat plane. This behavior indicates that the system evolution obeys, irrespective of time, the rule laid down by Eq.(2) i.e. a time-invariant, deterministic, nonlinear model for the ENSO slow manifold. Equation (2) forms the basis for prediction of $y_{i+2\tau}$ using previous 4 datapoints $\tau$ apart (Fig. 2A).

**Dynamical equation underlying the evolution of the ENSO slow manifold**

The remarkable confinement of system trajectory to a flat plane for a range of values of $\tau$ (Figs. 2B-E) suggests existence of a differential equation underlying the nonlinear model (Eq. 2). The five-point building block of Fig. 2(A) enables conversion of Eq. (2) to a differential equation centered on $\hat{y}_i$ (see Materials and Methods in SM). The final equation is

$$\left[\frac{d^2\hat{y}}{d\hat{t}^2}\right]_{\hat{y}_i}\left[\frac{d^3\hat{y}}{d\hat{t}^3}\right]_{\hat{y}_i} + \frac{P_2}{2P_1}\left\{\hat{y}_i\left[\frac{d^3\hat{y}}{d\hat{t}^3}\right]_{\hat{y}_i} - \left[\frac{d\hat{y}}{d\hat{t}}\right]_{\hat{y}_i}\left[\frac{d^2\hat{y}}{d\hat{t}^2}\right]_{\hat{y}_i}\right\} + \frac{P_3}{P_1}\hat{y}_i\left[\frac{d\hat{y}}{d\hat{t}}\right]_{\hat{y}_i} = 0 \qquad (3)$$

Equation (3) is a third-order, nonlinear, homogeneous ODE and can be solved numerically as an initial value problem. Supplementary Material (Figs. S2 and S3) gives a detailed account of the solutions to Eq. (3) with different initial conditions and also with temporal variations of the coefficients. The two main outcomes are: (a) with coefficients obtained from the complete ENSO slow manifold data, Eq. (3) admits a self-sustained, quasi-periodic solution with period of 4.5 years which is in excellent agreement with the dominant periodicity of the ENSO slow manifold and (b) data shows that the coefficients in Eq. (3) can vary with time and this would excite decadal and multidecadal modes in the solution explaining the observed broad spectrum of the ENSO slow manifold. Moreover it is found that the terms in Eq. (3) can be identified to play distinct roles consistent with the known mechanism of the ENSO - the first two terms essentially represent an instability while the third term represents the negative feedback necessary to restrict the amplitude and introduce periodicity into the solution (see Supplementary Materials) somewhat akin to the delayed oscillator model (*32*).

**A linear predictive model for the ENSO slow manifold**

While the nonlinear model described in the preceding is the focus of the present paper, a linear model with the functional form

$$\hat{y}_{i+\tau} = A + B\hat{y}_{i-\tau} + C\hat{y}_i, \qquad (4)$$

is also found to describe the evolution of the ENSO slow manifold, at short time delays, reasonably well (Supplementary Materials, Fig. S4). The linear model (Eq. 4) uses embedding in three dimensions ($m = 3$) and its state point building block comprises of the middle three points of Fig. 2(A). Equation (4) can be used for prediction of $\hat{y}_{i+\tau}$ using previous 2 datapoints $\tau$ apart. Although less-accurate and less-skillful than its nonlinear counterpart, the linear model serves two important purposes. First, it suggests why linear models with stochastic forcing are able to produce reasonably skillful ENSO forecasts. Second, it provides the necessary contrast between linearity and nonlinearity underscoring the importance of capturing the correct nonlinear behavior towards improved prediction skill and weakened seasonality of predictions (SPB). The linear description of system evolution degrades quickly with increase of the embedding time delay $\tau$ (Supplementary Materials, Fig. S4) consistent with the fact that Eq. (4), unlike Eq. (2), cannot be rearranged to derive an underlying (linear) ODE centered on $\hat{y}_i$.

**Hindcasts**

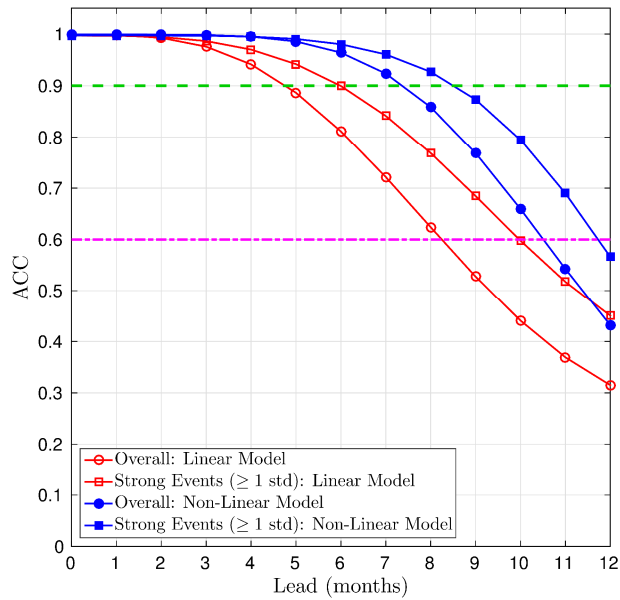

**Fig. 3. Hindcast prediction skill of the present nonlinear and linear models.** Anomaly Correlation Coefficient (ACC) of $y$, between predicted and original datapoints, is plotted against lead time. Circles show the results for all datapoints and squares show results only for the strong events i.e. values of $y$ exceeding one standard deviation. For reference, horizontal dashed line shows ACC = 0.9 and dashed-dotted line shows ACC = 0.6.

Nonlinear and linear model hindcasts for the ENSO slow manifold are generated using Eqs. (2) and (4) respectively. The first 30% points of the ENSO slow manifold timeseries are used for training the models and hindcasts, at different leads, are generated for the remaining 70% datapoints (see Materials and Methods in SM); sample hindcast timeseries are shown in the Supplementary Materials (Figs. S5 and S6). Figure 3 shows hindcast skill of both models with Anomaly Correlation coefficient (ACC) plotted against prediction lead time. The nonlinear model dramatically outperforms the linear model everywhere exhibiting very high skill (ACC > 0.9) at lead times up to 7 months; for linear model the corresponding lead time is 4.5 months. Multi-model ensemble (MME) hindcasts, that are considered to be the most skillful compared to individual coupled model hindcasts, yield ACC of 0.86 for Niño3.4 SST (an ocean-based index for monitoring ENSO) at lead time of 6 months (*33*). In comparison, the present nonlinear model yields a remarkably high ACC of 0.96 at lead time of 6 months for the ENSO slow manifold at practically negligible computational cost; the corresponding ACC for the linear model is 0.8. Useful hindcasts (ACC > 0.6) are possible with the nonlinear(linear) model up to lead time of 10.5(8) months. However, we note that the drop in skill for both models at longer lead times is rather rapid compared to typical MME hindcasts. Figure 3 also shows that stronger events (>1 standard deviation) are better predicted by each model similar to the existing coupled model hindcasts (*33*); again nonlinear model performance is superior to that of the linear model.

Figure 4 examines the seasonality of hindcasts produced by both the models. Contours of ACC and RMSE (root-mean-squared error between the predicted and the original values) for hindcasts generated by the present linear (Figs. 4A,B) and nonlinear (Figs. 4C,D) models are plotted on a plane defined by the lead time and the prediction starting month (*9,34,35*). Quick loss of prediction skill and rapid error growth with the lead time around the starting month of April (Figs. 4A-D) is the signature of the well-known Spring Predictability Barrier (SPB) whose origin (*35-38*), however, still remains a subject of debate. Figure 4 shows that the SPB is dramatically weakened in the case of nonlinear model compared to the linear model hindcasts. This is brought out clearly in Figs. 4(E,F) where the most dramatic improvement(reduction) in ACC(RMSE), after switchover from linear to nonlinear model, occurs around a lead time of 9(7) months and the starting Calendar month of April(May) i.e. in the SPB zone. This shows that a major part of SPB is very likely related to the lack of correct modeling of the deterministic nonlinearity in the ENSO system and could explain success of some improved intermediate model initializations

(*39*) in alleviating the SPB. It is notable that the SPB is much milder in our nonlinear model (Figs.4 C,D) with useful skill up to a lead of 10.5 months, as compared to the dynamical or linear stochastic models where the useful skill is typically limited to about 6 to 8 months (*40,35*). Figures 3 and 4 forcefully bring out the importance of accounting for the deterministic nonlinearity towards skillful ENSO predictions at longer lead times.

**Discussion and Challenges**

The fact that our nonlinear dynamical model (Eq. 2) works well for two other indices of the ENSO (Supplementary Materials, Figs. S7 to S14) indicates that it is a robust representation of the observed ENSO nonlinearity. While the hindcasts for the coupled index MEI are very similar to those presented for the MEI.ext, the SPB in the SST-based Niño3.4 hindcasts is found to be stronger than in the MEI.ext and MEI predictions. This suggests that the contribution of the coupled ocean-atmosphere interaction to the overall nonlinearity is critical for reducing the SPB in predictions in addition to the nonlinearity contributed by the ocean alone.

While robust predictive potential of the nonlinear model is demonstrated here, real time predictions remain a challenge. It turns out that the climate noise, that was filtered out for the present analysis of the ENSO slow manifold, corrupts the present model significantly. Filtering when used in the real-time forecast mode introduces an error in the initial conditions that grows and deteriorates the forecasts quickly. This end-point problem is well known in time series forecasting and many solutions have been proposed with varied degree of success (*41*). We are exploring ways by which the slow manifold could be brought out without involving filtering, like the real-time monitoring index of the Madden Julian Oscillation (MJO, *42*), and also the initial error at the end point is reduced. We hope to report these findings in the near future.

Our methodology to discover the nonlinear model may have wide ranging applications in finding similar nonlinear models for other time series of geophysical phenomena such as the MJO, the Boreal Summer Intra-Seasonal Oscillations (BSISO), the Sunspots etc. as well as in areas other than climate science. Our preliminary tests indicate that the model works quite well for the Sunspot index time series; that will be reported elsewhere later. We plan to explore the generic nature of this nonlinear model in representing wide ranging geophysical systems.

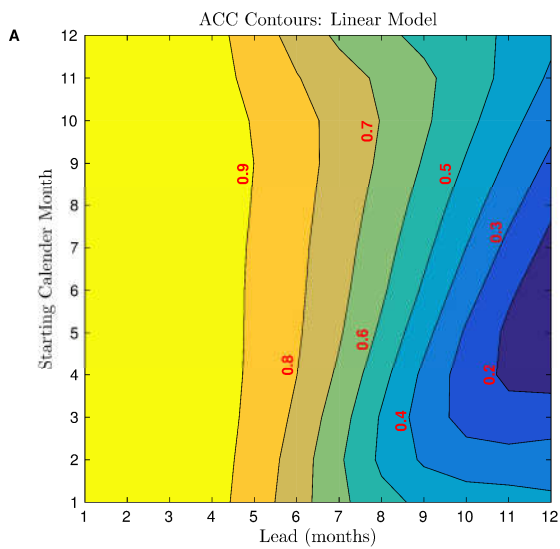
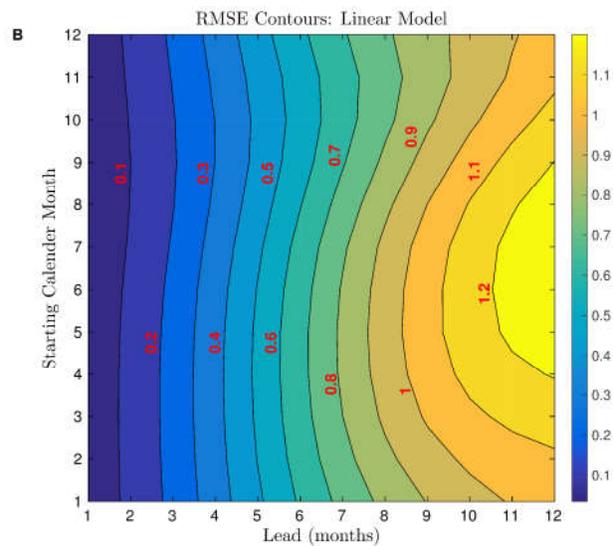
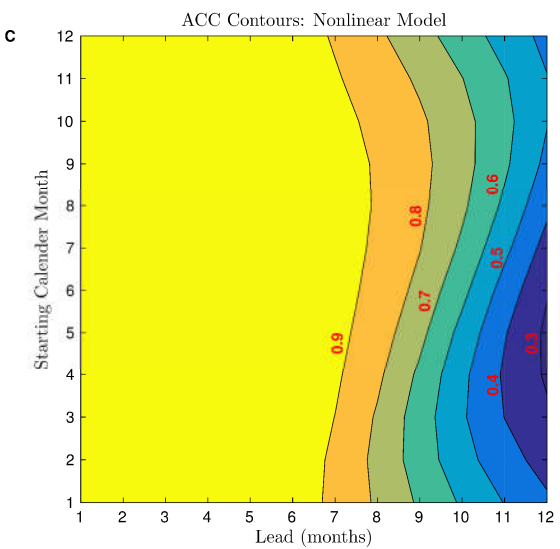
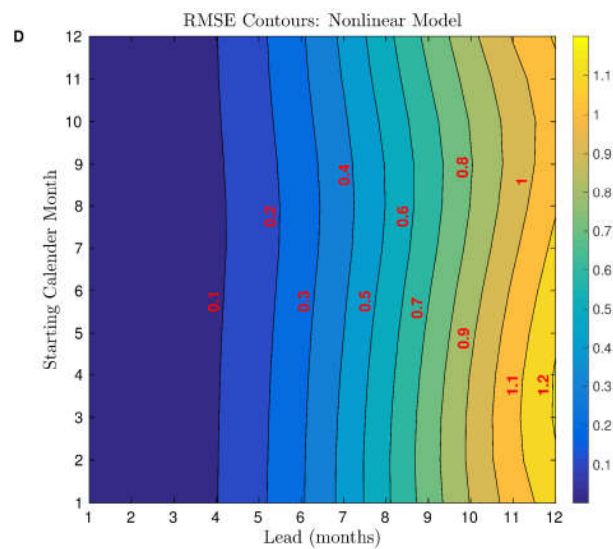
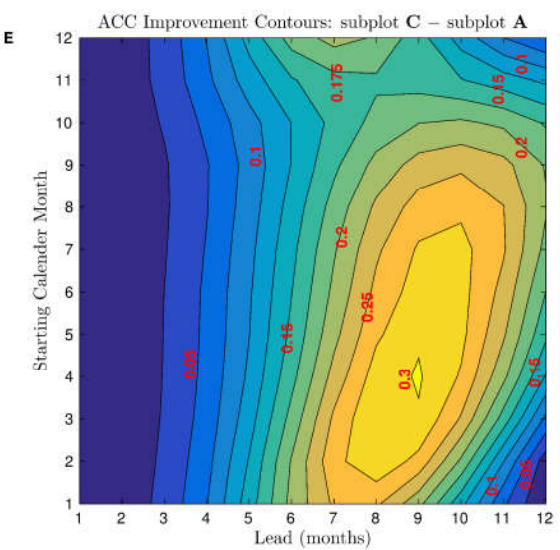
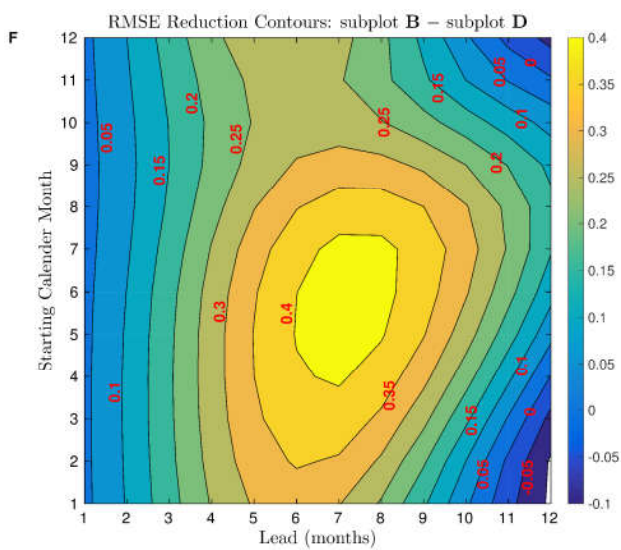

**Fig. 4. Seasonality of the hindcasts generated by the present linear and nonlinear models.** Contours of the Anomaly Correlation Coefficient (ACC) and Root-Mean-Squared-Error (RMSE) are plotted on plane defined by the lead time and the starting calendar month for the prediction. Faster drop(rise), with lead time, in the ACC(RMSE) around April (starting month = 4) indicates the well-known Spring Predictability Barrier (SPB) in **A-D**. Contours of the ACC improvement and RMSE reduction in nonlinear model hindcasts, as compared to those by the linear model, are shown in **E** and **F**.

**Acknowledgments:**

S.A.D. gratefully acknowledges the support of the Director, Indian Institute of Tropical Meteorology (IITM), Pune. IITM is an autonomous research institute under the Ministry of Earth Sciences (MoES), Government of India and is fully funded by MoES. S.A.D. sincerely thanks

Prof. G. Ambika (IISER, Pune) for suggesting the possibility of a differential equation underlying our nonlinear model, Mr. Rahul Chopde (Scientist, ARDE, Pune) for suggestion of using a DC shift in the model to avoid division by zero and Profs. O. N. Ramesh (IISc, Bangalore) and R. Narasimha (JNCASR, Bangalore) for insightful discussions. B.N.G. acknowledges the INSA for support through INSA Senior Scientist fellowship and CU for facilities during the course of this work.

S.A.D. and B.N.G. both conceived the study and wrote the manuscript. S.A.D. conceived of the idea of the nonlinear model for the ENSO and worked out the analytical and numerical details. B.N.G. suggested the use of nonlinear model for hindcast predictions, seasonality etc.

Data used in the analysis presented here are available at the following links with NOAA:

MEI.ext timeseries from NOAA-ESRL (https://www.esrl.noaa.gov/psd/enso/mei.ext/)

MEI timeseries from NOAA-ESRL (https://www.esrl.noaa.gov/psd/enso/mei/)

Niño3.4 timeseries from NOAA-CPC (http://www.cpc.ncep.noaa.gov/data/indices/sstoi.indices)

**Supplementary Materials (submitted as a separate file):**

Materials and Methods

Supplementary Text

Figs. S1-S15

# Supplementary Materials for

Discovery of a Phenomenological Dynamical Model for Predicting the El Niño-Southern Oscillation

Shivsai Ajit Dixit[1*], B. N. Goswami[2]

Correspondence to: sadixit@tropmet.res.in

**This PDF file includes:**

Materials and Methods
Supplementary Text
Figs.S1 to S15

**Materials and Methods**

MEI.ext index of the ENSO and the ENSO slow manifold

The MEI.ext (Extended Multivariate ENSO Index) is derived (*20*) as the first non-rotated principal component (PC) of sea-level pressure (SLP, HadSLP2 dataset) and sea-surface temperature (SST, HadSST2 dataset) fields in the Pacific Ocean (30N–30S and 100E–70W). These two variables respectively represent variability in the atmosphere and the ocean and therefore MEI.ext signifies the coupled character of the ENSO. MEI.ext has been compared against several other ENSO monitoring indices (*20*) and has been found to be quite robust. MEI.ext spans a long period of 135 years (January 1871 to December 2005) giving a timeseries of 1620 datapoints (12 datapoints per year). Nonzero skewness factor of MEI.ext ($S_{\tilde{y}} = 0.322$, see Fig. 1A) suggests the well-known El Niño-La Niña asymmetry and indicates the nonlinearity of ENSO (*10,11*) captured by MEI.ext. Positive values of MEI.ext correspond to El Niño events and negative values correspond to La Niña events.

The 'slow manifold' ($y$ in Fig. 1A) is obtained by applying a spectral (FFT-based) filter to the MEI.ext timeseries ($\tilde{y}$ in Fig. 1A), to remove harmonic components (climate noise) with periods less than and equal to one year; this corresponds to a low-pass cutoff frequency of 0.0833 per month (see Fig. 1B). The procedure involves computing FFT of $\tilde{y}$, setting the Fourier coefficients of harmonic components above the cutoff frequency to zero and then inverting the transformation to obtain $y$. Figure 1(B) shows that $y$ retains most of the energy (about 95%) of $\tilde{y}$ - the energy of a signal (with zero mean) is, by definition, its variance and is given by the area under the curve of its power spectral density (PSD).

Nonlinear model construction

As discussed in the main text of the paper, as the ENSO slow manifold dynamics unfolds in time, it is possible that a certain functional relationship between the coordinates of the state point in the five-dimensional phase space holds irrespective of time. For the building block of Fig. 2(A), such a time-invariant functional relationship, in its most general form, will be

$$f(y_{i+2\tau}, y_{i+\tau}, y_i, y_{i-\tau}, y_{i-2\tau}) = 0.  \quad (S1)$$

With the expectation of $f$ being a deterministic and nonlinear function, one may write a power-series approximation to Eq. (S1) as

$$a_0 + \sum_{m=i-2\tau}^{m=i+2\tau} a_m y_m + \sum_{m=i-2\tau}^{m=i+2\tau} \sum_{n=i-2\tau}^{m=i+2\tau} b_{mn} y_m y_n + \cdots = 0, \quad (S2)$$

where the first summation contains linear terms, second contains quadratic terms and so on and the coefficients $a_0$, $a_m$, $b_{mn}$ etc. are, in general, functions of $\tau$. Note that $y_{i+2\tau}$ in Eq. (S2) deterministically depends only on the values of $y$ from the past. Equation (S2) is in fact akin to the generalized Volterra series description which is commonly used to describe time-invariant relationships between input and output of a nonlinear dynamical system (*31*).

It is not immediately clear which terms in Eq.(S2) are important and need to be retained. However, we are guided by the following three simple considerations:

(a) If one wishes to obtain an unambiguous prediction of $y_{i+2\tau}$ using Eq. (S2), the terms nonlinear in $y_{i+2\tau}$ need to be left out.

(b) If one rearranges Eq. (S2), for prediction purposes, such that $y_{i+2\tau}$ is on the left side and everything else is moved to the right side, denominator on the right side should not tend to or become zero.

(c) The number of terms (except for $a_0$) may be restricted to three, in order to facilitate "visualization" (by three-dimensional plotting) of the extent to which Eq. (S2) is getting satisfied.

Consideration (b) would cause the prediction of $y_{i+2\tau}$ to blow off to unrealistically high or low values and presents a major hurdle in predictions. However, it turns out that this issue is easily addressed by adding a simple DC shift *S* to the *y* signal; Fig. S1 in Supplementary Materials shows the dependence of the hindcast skill of the present nonlinear model (Eq. 2) on the DC shift *S*. Thus, we need to cast the entire problem in terms of $\hat{y} = y + S$, make predictions of $\hat{y}$ and then recover the prediction for *y* by subtracting *S* from it. Therefore in terms of $\hat{y}$, Eqs. (S1) and (S2) become

$$F(\hat{y}_{i+2\tau}, \hat{y}_{i+\tau}, \hat{y}_i, \hat{y}_{i-\tau}, \hat{y}_{i-2\tau}) = 0, \tag{S3}$$

$$A_0 + \sum_{m=i-2\tau}^{m=i+2\tau} A_m \hat{y}_m + \sum_{m=i-2\tau}^{m=i+2\tau} \sum_{n=i-2\tau}^{m=i+2\tau} B_{mn} \hat{y}_m \hat{y}_n + \cdots = 0. \tag{S4}$$

Guided by the considerations (a) and (c) mentioned above, different combinations were tested for satisfaction of Eq. (S4) using different delay times ranging from $\tau = 1$ to 3 months covering intervals of 4 to 12 months on the time axis (Fig. 2A). After several trials involving three-dimensional plotting, it was discovered that the combination that satisfies Eq. (S4) to an excellent approximation, is given by

$$A_0 + B_{ii} \hat{y}_i \hat{y}_i + B_{i-\tau, i+\tau} \hat{y}_{i-\tau} \hat{y}_{i+\tau} + B_{i-2\tau, i+2\tau} \hat{y}_{i-2\tau} \hat{y}_{i+2\tau} = 0, \tag{S5}$$

which may be rearranged to yield the present nonlinear model (Eq. 2 of the main text) wherein $A = -A_0 / B_{ii}$, $B = -B_{i-2\tau, i+2\tau} / B_{ii}$ and $C = -B_{i-\tau, i+\tau} / B_{ii}$.

Derivation of the differential equation from the nonlinear model

We now discuss how the differential equation centered on $\hat{y}_i$ (Eq. 3) is constructed from the nonlinear model of Eq. (2). Towards this, we note that the first- and the second-order derivatives at $\hat{y}_i$ - using the central-difference approximation (Taylor series) with dimensionless step size $\hat{\tau}$ - are

$$\left[\frac{d\hat{y}}{d\hat{t}}\right]_{\hat{y}_i} = \frac{\hat{y}_{i+\tau} - \hat{y}_{i-\tau}}{2\hat{\tau}}, \tag{S6}$$

$$\left[\frac{d^2\hat{y}}{d\hat{t}^2}\right]_{\hat{y}_i} = \frac{\hat{y}_{i+\tau} - 2\hat{y}_i + \hat{y}_{i-\tau}}{\hat{\tau}^2}. \tag{S7}$$

Rearranging Eqs. (S6) and (S7) yields

$$\hat{y}_{i+\tau} - \hat{y}_{i-\tau} = 2\hat{\tau}\left[\frac{d\hat{y}}{d\hat{t}}\right]_{\hat{y}_i}, \tag{S8}$$

$$\hat{y}_{i+\tau} + \hat{y}_{i-\tau} = 2\hat{y}_i + \hat{\tau}^2\left[\frac{d^2\hat{y}}{d\hat{t}^2}\right]_{\hat{y}_i}. \tag{S9}$$

Squaring and subtracting Eqs. (S8) and (S9) gives

$$\hat{y}_{i+\tau}\hat{y}_{i-\tau} = \hat{y}_i^2 + \frac{\hat{\tau}^4}{4}\left[\frac{d^2\hat{y}}{d\hat{t}^2}\right]_{\hat{y}_i}^2 + \hat{\tau}^2\hat{y}_i\left[\frac{d^2\hat{y}}{d\hat{t}^2}\right]_{\hat{y}_i} - \hat{\tau}^2\left[\frac{d\hat{y}}{d\hat{t}}\right]_{\hat{y}_i}^2. \tag{S10}$$

Similarly, with the central difference approximations having dimensionless step size $2\hat{\tau}$, one obtains

$$\hat{y}_{i+2\tau}\hat{y}_{i-2\tau} = \hat{y}_i^2 + 4\hat{\tau}^4\left[\frac{d^2\hat{y}}{d\hat{t}^2}\right]_{\hat{y}_i}^2 + 4\hat{\tau}^2\hat{y}_i\left[\frac{d^2\hat{y}}{d\hat{t}^2}\right]_{\hat{y}_i} - 4\hat{\tau}^2\left[\frac{d\hat{y}}{d\hat{t}}\right]_{\hat{y}_i}^2. \tag{S11}$$

Substituting Eqs. (S10) and (S11) in the nonlinear model of Eq. (2) and simplifying yields

$$P_1\left[\frac{d^2\hat{y}}{d\hat{t}^2}\right]_{\hat{y}_i}^2 + P_2\left\{\hat{y}_i\left[\frac{d^2\hat{y}}{d\hat{t}^2}\right]_{\hat{y}_i} - \left[\frac{d\hat{y}}{d\hat{t}}\right]_{\hat{y}_i}^2\right\} + P_3\hat{y}_i^2 + P_4 = 0. \tag{S12}$$

Here $P_1$ through $P_4$ are coefficients and are functions of $\hat{\tau}$

$$P_1 = \left(4B + \frac{C}{4}\right)\hat{\tau}^4,$$
$$P_2 = (4B + C)\hat{\tau}^2,$$
$$P_3 = B + C - 1,$$
$$P_4 = A,$$

wherein $A$, $B$ and $C$ (see Eq. 2) in turn depend on $\hat{\tau}$.

Equation (S12) represents the underlying differential equation that governs the evolution of the slow manifold of ENSO. However, (S12) is non-homogeneous and one more differentiation with respect to $\hat{t}$ yields the final differential equation (see Eq. 3). Note that Eq. (3) holds only if the Taylor series approximation for the derivatives is valid. This is so when the largest step size chosen for the finite-difference approximations is small enough i.e. $2\hat{\tau} < 1$ or $\hat{\tau} < 0.5$. This is consistent with the observation that the ENSO slow manifold data remains confined to a flat plane up to $\hat{\tau} = 0.43$ (see Fig. 2).

Methodology for generating linear and nonlinear model hindcasts

Hindcasts at longer lead times can be generated from linear and nonlinear models (Eqs. 2 and 4) by the following two strategies:
(a) Dilating the building block of Fig. 2(A) i.e. using larger values of delay $\tau$ or
(b) Translating the building block with the smallest delay ($\tau = 1$) forward along the timeline through the short-lead predictions.

Strategy (a) involves direct leap-frogging to the long-lead forecast. Strategy (b) produces a one-month-lead forecast first. The building block (Fig. 2A) then slides forward by one point and makes the next one-month-lead forecast that uses the forecasted value at the earlier point. This process repeats till the required long-lead forecast is achieved. It is found that the strategy (b) yields more skillful predictions mainly because the trajectory does not remain confined to a flat plane for large delays (see Fig. 2). We use strategy (b) for our hindcasts. The model under consideration (linear or nonlinear) is trained using the first 30% i.e. 486 datapoints (40.5 years of data) of the ENSO slow manifold; this gives values of coefficients $A$, $B$ and $C$ in the model (Eq. 2 or 4). The trained model is then used to obtain hindcasts for the remaining 70% of the timeseries (94.5 years) at different lead times. The ACC between the set of predicted datapoints and original slow manifold values is then computed at each lead time. The dependence of ACC on model training period is assessed separately in the Supplementary Materials (Fig. S15). This assessment shows that the results presented in Fig. 3 of the main text are quite robust.

**Supplementary Text**

Dependence of the linear and nonlinear model hindcast skill on the DC shift $S$

Results presented in the main text of the paper use hindcasts generated with linear and nonlinear models (Eqs. 4 and 2 respectively) that use DC shift $S = 20$. Here we assess the effect of varying the DC shift on the hindcast skill of our models as shown in Fig. S1. For the linear model (Fig. S1A), DC shift has no effect on the model skill. This is so because the model itself is linear. For the nonlinear model (Fig. S1B), however, the model skill strongly depends on the DC shift, due to model nonlinearity, and improves with the DC shift. It may be noted that for $S = 0$, the model skill is very low and improves dramatically for $S = 5$ after which the improvement slows down and for $S = 20$, the skill is almost saturated. This justifies the DC shift $S = 20$ used in the main text of the paper for both the models.

Numerical solutions of differential equation (Eq. 3 of the main text)

Equation (3) of the main paper may be solved numerically as an initial value problem. In what follows, we discuss various aspects of solutions to Eq. (3).

For embedding time delay $\tau = 1$ month i.e. $\hat{\tau} = 0.14$ (see Fig. 2B), fitting the nonlinear model (Eq. 2) to the complete ENSO slow manifold data yields the model coefficients $A = 0.1722$, $B = -0.3539$ and $C = 1.353$. Using these, one obtains $P_1 = -4.5 \times 10^{-4}$, $P_2 = -1.28 \times 10^{-3}$ and $P_3 = -9 \times 10^{-4}$ (see Materials and Methods). The initial values of $\hat{y}$, $d\hat{y}/d\hat{t}$ and $d^2\hat{y}/d\hat{t}^2$ for solving Eq. (3) may be taken from a point on the ENSO slow manifold; differentiations of $\hat{y}$ may be obtained by central difference approximation. We now analyze the solutions to Eq. (3) in two ways, first, by using the coefficient values $P_2/2P_1 = 1.4113$ and $P_3/P_1 = 0.9433$ obtained from the complete ENSO slow manifold data and next, by allowing $P_2/2P_1$ and $P_3/P_1$ to vary with time.

Figure S2(A) shows the solution to Eq. (3) obtained using ode45 solver of MATLAB subject to a representative initial condition corresponding to $t = 1499$ months on the ENSO slow manifold. Self-sustained near-periodic oscillations with the dimensionless time period $\hat{T}_P = T_P/T = 7.8$ are evident. With $T = 7$ months, the integral timescale of the ENSO slow manifold (see Fig. 1C), this translates to $T_P = 54.6$ months i.e. 4.55 years periodicity which agrees remarkably well with the dominant period in the ENSO slow manifold spectrum (Fig. 1B). That the solution is non-sinusoidal is clear from "pinched" troughs and "broadened" crests in the second derivative curve (Fig. S2A). It is easy to see that substituting a sinusoid as solution in Eq. (3) leads to a characteristic equation which is satisfied only if the sinusoid frequency is complex-valued indicating that a pure sinusoid cannot be a solution to Eq. (3). Solutions with randomly chosen initial conditions from the ENSO slow manifold can be obtained. From such 50,000 random choices, the probability density function (PDF) and the cumulative distribution function (CDF) of the resulting time period $T_P$ have been constructed. Figure S2(B) shows that the most probable period is about 54 months where CDF crosses 50% mark. While the amplitude of the solution depends on the initial condition (not shown), Fig. S2(B) shows that any initial condition from the ENSO slow manifold always results in the solution period that is close to the dominant period of ENSO. This demonstrates the robustness of the differential equation (Eq. 3) under consideration.

In order to gain insight into the roles of various terms in Eq. (3), we investigate the effect of relative strengths of these terms. Strengths of the second and third terms with respect to the first term may be altered by changing the values of coefficients $P_2/2P_1$ and $P_3/P_1$ respectively. It is observed that this has a distinct effect on the periodicity of the solution. Figure S2(C) shows the contour plots of period $T_P$ of the solution on $P_2/2P_1$ versus $P_3/P_1$ plane; $t = 1499$ months is used as the initial condition for this plot. The point corresponding to the values of ratios obtained from the complete ENSO slow manifold is shown by a big cross. Clearly, Eq. (3) admits solutions with varying periodicities ranging from yearly to decadal and multidecadal variability. Figure S2(C) shows that for a given strength of the third term (i.e. a given value of $P_3/P_1$), increasing the strength of the second term (i.e. the value of $P_2/2P_1$) results in increased period of oscillations towards decadal and multidecadal modes. On the contrary for a given strength of the second term, an increase in the strength of the third term reduces the period of oscillations. Figure S2(D) shows the limit cycles, corresponding to different values of coefficients $P_2/2P_1$ and $P_3/P_1$, associated with the solutions with $t = 1499$ months as the initial condition. The shapes of all limit cycles are clearly asymmetric indicating non-sinusoidal periodic solutions.

The third term appears to be crucial for obtaining an oscillating solution. If this term is not present ($P_3/P_1 = 0$), the solution is found to blow off exponentially to very high values (Fig. S2E). The exponential blow off is less rapid if only the first term exists ($P_2/2P_1 = 0$ and $P_3/P_1 = 0$) and starts becoming progressively more rapid as the second term is included and increased in strength ($P_2/2P_1 > 0$ and $P_3/P_1 = 0$, Fig. S2E). Further, if $P_3/P_1 < 0$, then also the solution shows exponential blow off with the magnitude of the blow off reducing as $P_3/P_1$ tends to zero from the negative side (not shown). Furthermore, in comparison to the other terms if the strength of the third term is increased ($P_3/P_1 \gg 1$ and $P_3/P_1 \gg P_2/2P_1$), the solution period shortens and the amplitude diminishes in a progressive fashion (Fig. S2F). Thus, the first two terms in Eq. (3) essentially represent an instability while the third term may be interpreted as modeling the negative feedback (in terms of *local* quantities) that keeps the amplitude in check and introduces essential periodicity in the solution.

It is also observed (not shown) that if both the second and the third terms in Eq. (3) are much stronger than the first term ($P_3/P_1 \gg 1$ and $P_2/2P_1 \gg 1$) - which is equivalent to the first term being absent - then the solution is a sinusoid with period $T_P = 2\pi T / \sqrt{2P_3/P_2}$, where $T$ = 7 months is the integral timescale of the ENSO slow manifold (see Fig. 1C); this may be easily checked by dropping the first term of Eq. (3) and then substituting a sinusoid into it.

We now investigate the variation of $P_2/2P_1$ and $P_3/P_1$ with time from the ENSO slow manifold data. Ratios $P_2/2P_1$ and $P_3/P_1$ in Eq. (3) of the main paper could, in general, be functions of time and this opens up a possibility of exciting different modes of variability as suggested by Fig. S2(C). Towards this end, we evaluate these ratios by fitting the nonlinear model of Eq. (2) to the slow manifold datapoints belonging to a window of size $L_{win}$. The values so obtained are assigned to the time coordinate corresponding to the centre of the window. The window is then made to progressively slide forward along the time axis giving $P_2/2P_1$ and $P_3/P_1$ values as functions of time.

Figures S3(A,B) respectively show the contours of the ratios $P_2/2P_1$ and $P_3/P_1$ plotted on a plane defined by the time coordinate and the window size. The smallest window size used is $L_{win} = 47$ months corresponding to the dominant ENSO periodicity of 4 years and the largest window size is 851 months i.e. approximately 71 years. Both plots show that for shorter windows ($L_{win} \leq 87$ months i.e. 7.4 years), the ratios undergo significant variations with time and for longer windows they show only mild variations. The horizontal dashed line in Figs. S3(A,B) shows $L_{win} = 87$ months as an approximate demarcation between strong and mild temporal variations of both the ratios. Figures S3(C,E) show the normalized probability density functions (PDFs) of $P_2/2P_1$ and $P_3/P_1$ for the window size ranges $47 \leq L_{win} \leq 851$ (full range) and $47 \leq L_{win} \leq 87$ (range of short window sizes) respectively. Also shown therein is the PDF of normal distribution. It is clear that the PDFs in either case depart significantly from Gaussianity.

Figures S3(D,F) respectively show the joint probability density contours for the two window size ranges under consideration in Figs. S3(C,E). Dashed-dotted lines in these plots indicate the values $P_2/2P_1 = 1.4113$ and $P_3/P_1 = 0.9433$ obtained from the complete ENSO slow manifold data. Figure S3(D) shows that for the full range of window sizes, the joint probability density is dominated by the near-constancy in the values of ratios for larger window sizes causing the contours to be tightly confined around $P_2/2P_1 = 1.4113$ and $P_3/P_1 = 0.9433$. For the range of shorter window sizes however, Fig.S3(F) shows that the contours are well spread out with many different

combinations of the ratio values showing dense probabilities of occurrence. In particular, combinations with $P_3/P_1 < 0.9433$ may be expected to excite decadal and multidecadal variabilities as suggested by Fig. S2(C).

Formulation of a linear predictive model

Following the same line of arguments as for the formulation of the nonlinear model (see Materials and Methods), one can construct a time-invariant, deterministic, linear model given by Eq. (4) in the main paper. In order to see how well it describes the ENSO slow manifold evolution, Fig. S4 plots the phase space trajectory of the ENSO slow manifold (after DC-shift i.e. $\hat{y} = y + 20$) in the three-dimensional phase space having coordinates $\hat{y}_i$, $\hat{y}_{i-\tau}$ and $\hat{y}_{i+\tau}$. Two delay times, $\tau = 1$ month and $\tau = 3$ months, are shown. It is clear that while the system trajectory remains confined to a flat plane for $\tau = 1$ month, it does not do so for $\tau = 3$ months. Figure S4 is to be compared with Fig. 2 of the main paper which shows similar trajectory evolution for the case of the nonlinear model (Eq. 2).

Sample timeseries for nonlinear and linear model hindcasts

Figures S5 and S6 respectively show the plots of sample hindcast timeseries using the present nonlinear and linear models (Eqs. 2 and 4). Procedure of generating these hindcasts is described in Materials and Methods. Figure captions provide the other relevant details.

Results for the ENSO index MEI (ocean + atmosphere based)

The MEI (Multivariate ENSO Index), a more comprehensive sibling index of the MEI.ext, is derived as the first non-rotated principal component (PC) of six fields - namely the sea-level pressure (SLP), sea-surface temperature (SST), zonal and meridional surface wind components, near-surface air temperature and total cloudiness (*43,44*) - from the COADS over the Pacific Ocean (30N–30S and 100E–70W). The MEI.ext used in the main text of the paper, on the other hand, uses only the SLP and SST fields. Another important difference between these two indices is the spatial clustering step used in the construction of MEI and omitted in the construction of MEI.ext (*20*). Similar to MEI.ext, MEI also signifies the coupled character of the ENSO and is updated every month. The MEI currently spans the period from January 1950 to June 2017 giving a timeseries of 810 datapoints (12 datapoints per year). Positive values of MEI correspond to El Niño events and negative values correspond to La Niña events. Figures S7-S10 show the results for MEI and are to be compared with Figs. 1-4 for the MEI.ext from the main text of the paper. For fair comparison, the spectral cutoff for the slow manifold, nonlinear and linear model training periods, model DC shift etc. are kept identical to those used while processing the MEI.ext.

Figure S7 shows the MEI timeseries and the corresponding ENSO slow manifold (Fig. S7A), the power spectral density (Fig. S7B) and the autocorrelation coefficient and integral timescale of the slow manifold (Fig. S7C). Figure S8 shows that the MEI slow manifold data also conform to our nonlinear model (Eq. 2 of the main text) for a range of values of time delays; Fig. S8(A-D) may be compared with Fig. 2(B-E) of the main text of the paper.

Figure S9 shows the comparison of hindcast skills of the linear and nonlinear models for the ENSO slow manifold derived from the MEI. Nonlinear(linear) model yields skillful hindcasts up to lead time of 10.5(7.5) months which is very similar to the results of Fig. 3 for the MEI.ext from the main text of the paper.

Figure S10 shows the comparison of the seasonality of linear and nonlinear model hindcasts for the ENSO slow manifold derived from the MEI. This should be compared with Fig. 4 for the MEI.ext from the main text. Broadly the conclusions are the same. Nonlinear model skill is far better than the linear model including the reduced seasonality. The nonlinear model has weakened SPB compared to the linear model (Fig. S10A-D); this is very similar to the MEI.ext result (Fig. 4A-D). The improvement contours of Fig. S10(E,F) show some differences compared to Fig. 4(E,F) in terms of the location of maximum improvement. These may be attributed to the differences in the methodologies used to construct the MEI and the MEI.ext. We believe that the spatial clustering step of MEI construction, omitted while constructing the MEI.ext, is responsible for this difference. However, we hasten to add that the main conclusion that the nonlinear model outperforms the linear model remains intact.

Results for the ENSO index Niño3.4 (ocean based)

The Niño3.4 Index is an ocean-based index of ENSO indicative of the sea-surface temperature (SST) in the Pacific Ocean (5N–5S and 170W–120W). Contrary to the MEI.ext and the MEI, Niño3.4 does not account for the coupled character of the ENSO. Weekly and monthly versions of Niño3.4 are available; here we use the monthly version of Niño3.4 which extends from January 1982 to June 2017 giving a timeseries of 426 datapoints (12 datapoints per year). To remove the annual cycle, monthly mean values are computed for the entire period and are subtracted from the corresponding monthly values to obtain Niño3.4 monthly anomalies. This anomaly timeseries is then subjected to our nonlinear and linear model analysis. Figures S11-S14 show results for Niño3.4 and these are to be compared with Figs. S7-S10 for the MEI in the preceding and Figs. 1-4 for the MEI.ext from the main text of the paper. While analyzing Niño3.4 data, the spectral cutoff for the slow manifold, nonlinear and linear model training periods, model DC shift etc. are kept identical to those used while processing the MEI and the MEI.ext.

Figure S11 shows the Niño3.4 monthly anomaly timeseries and the corresponding ENSO slow manifold (Fig. S11A), the power spectral density (Fig. S11B) and the autocorrelation coefficient of the slow manifold (Fig. S11C). Figure S12 shows that the Niño3.4 slow manifold data also conform to our nonlinear model (Eq. 2 of the main text) for a range of values of time delays.

Figure S13 shows the comparison of hindcast skills of the linear and nonlinear models for the ENSO slow manifold derived from Niño3.4. Nonlinear(linear) model yields skillful hindcasts up to lead time of 11(6.5) months which is similar to the results of Fig. 3 from the main text for the MEI.ext.

Figure S14 shows the comparison of the seasonality of linear and nonlinear model hindcasts for the ENSO slow manifold derived from the Niño3.4 which is an ocean-only index. Figure S14 should be compared with Fig. 4 of the main text as well as Fig. S10 that pertain to the ocean + atmosphere indices MEI.ext and MEI respectively. Some interesting differences may be noted. First of all, both linear and nonlinear models show degraded performance in Niño3.4 case (Fig. S14) compared to the MEI.ext (Fig. 4) and MEI (Fig. S10) cases. Next and more important is the observation that the nonlinear model shows stronger SPB in the Niño3.4 case (Fig. S14C) compared to the MEI.ext (Fig. 4C) and MEI (Fig. S10C) cases.The improvement contours of Fig. S15(E,F) also show differences compared to Fig. 4(E,F) in terms of the location of maximum improvement. These may be attributed to the ocean-only or ocean + atmosphere nature of indices under consideration. We note however that the conclusion that the nonlinear model outperforms the linear model remains intact even for Niño3.4 index.

These results suggest that the reflection, in an ENSO monitoring index, of the coupled ocean-atmosphere interaction which is at the very heart of ENSO, is critical to the reduction of SPB in addition to modeling the nonlinearity correctly. Alternatively the contribution of the ocean-atmosphere interaction to the overall nonlinearity appears to be significant and cannot be neglected in comparison to the ocean nonlinearity alone. From a comparative study of the results for the three ENSO indices namely the MEI.ext, MEI and Niño3.4, the MEI.ext emerges as the most robust index of ENSO with minimal SPB and enhanced nonlinear model skill. This makes a strong case for extending the MEI.ext beyond 2005 and updating it in realtime as its sibling index MEI.

Dependence of the linear and nonlinear model hindcast skill on the training period

In order to assess the effect on prediction skill of the training period of the linear and nonlinear models (Eqs. 4 and 2) of the main paper, we plot in Fig. S15(A,B), the Anomaly Correlation Coefficient (ACC) of the hindcast predictions with both models trained for different periods. It is clear that for the training period as low as 16 months, which is 0.01% of the total timeseries length of 1620 months, the skill of the linear model is unacceptably low (Fig. S15A) as compared to that of the nonlinear model (Fig. S15B). This is an indication that the nonlinear model better captures the underlying dynamics of the ENSO slow manifold. As the training period increases, the skill for both models shows marked improvement. For the training period of 486 months, which is 30% of 1620 months, used in the main paper, the skill of both models is saturated as no further improvement in the skill is seen when the training period is increased to 810 months (i.e. 50% of 1620 months).

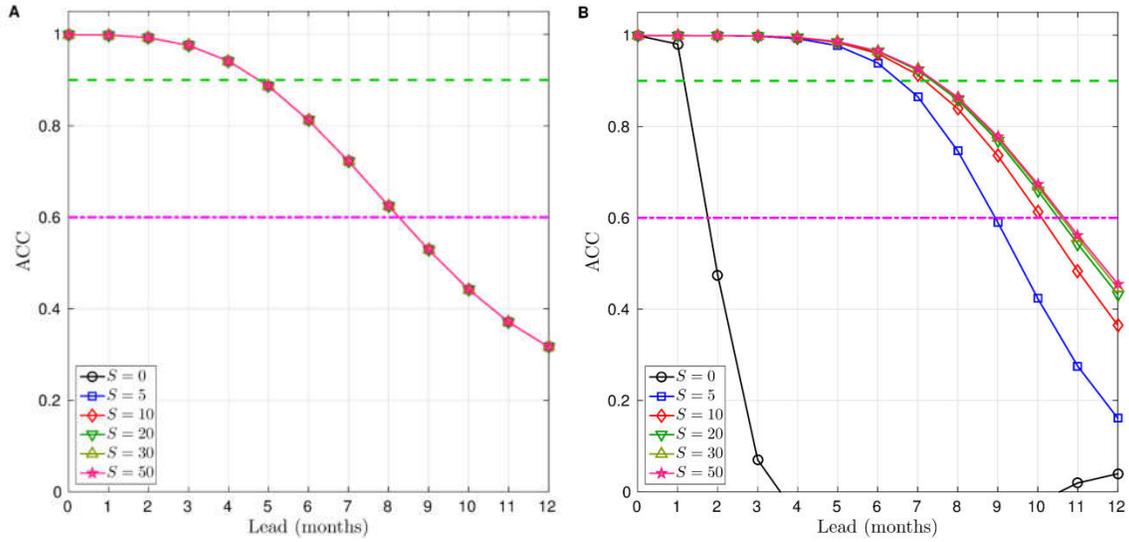

**Fig. S1. Effect of the DC shift *S*, used in the linear and nonlinear models, on the hindcast prediction skill.** Anomaly Correlation Coefficient (ACC) of $y$, between predicted and original datapoints is plotted against lead time for **A,** linear model (Eq. 4) and **B,** nonlinear model (Eq. 2). Curves for different DC shifts are shown. For reference, horizontal dashed line shows ACC = 0.9 and dashed-dotted line shows ACC = 0.6.

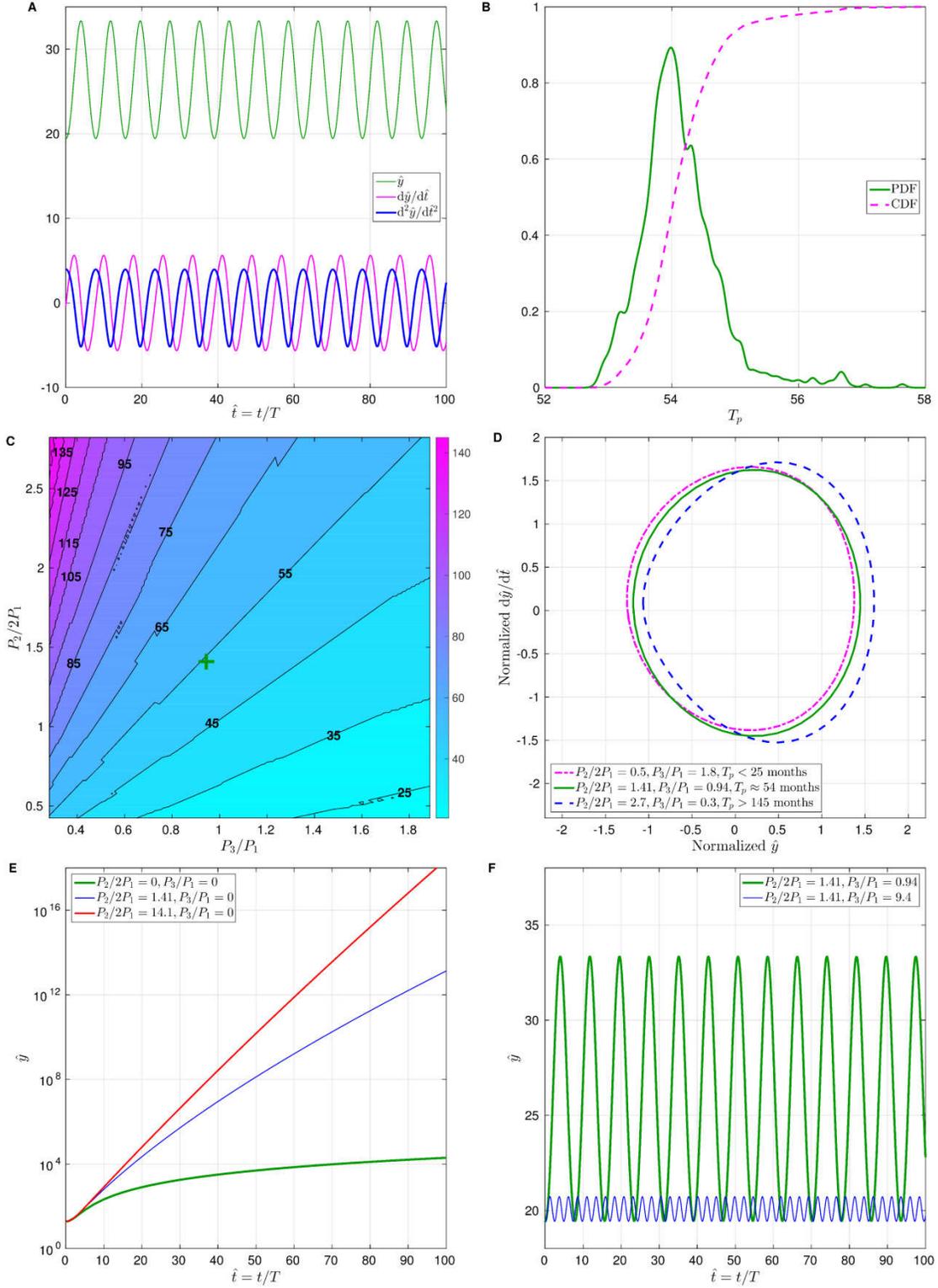

**Fig. S2. Numerical solutions to the differential equation (Eq. 3). A,** Numerical solution to Eq. (3) subject to a set of initial conditions at $t = 1499$ months taken from the ENSO slow manifold. **B,** Probability density function (PDF) and Cumulative Distribution

Function (CDF) of solution period $T_P$ generated by solving Eq. (3) subject to 50,000 initial conditions selected randomly from the ENSO slow manifold. **C,** Contours of $T_P$ for variation in coefficients $P_2/2P_1$ and $P_3/P_1$ of Eq. (3). The combination of $P_2/2P_1$ and $P_3/P_1$ values arising from the complete slow manifold data is shown by the big cross. **D,** Limit cycle shapes for values of $P_2/2P_1$ and $P_3/P_1$ corresponding to the lower right and upper left corners and the big cross of plot **C**. All limit cycles have asymmetric shapes indicating non-sinusoidal solution. **E,** Exponentially growing solutions (instability) to Eq. (3) with only the first two terms retained ($P_3/P_1 = 0$) and for different strengths of the second term (relative to the first term). **F,** Effect of the strength of the third term on period and amplitude of oscillating solutions to Eq. (3). For plots **C-F**, initial conditions from the slow manifold at $t = 1499$ months are used for obtaining the solutions.

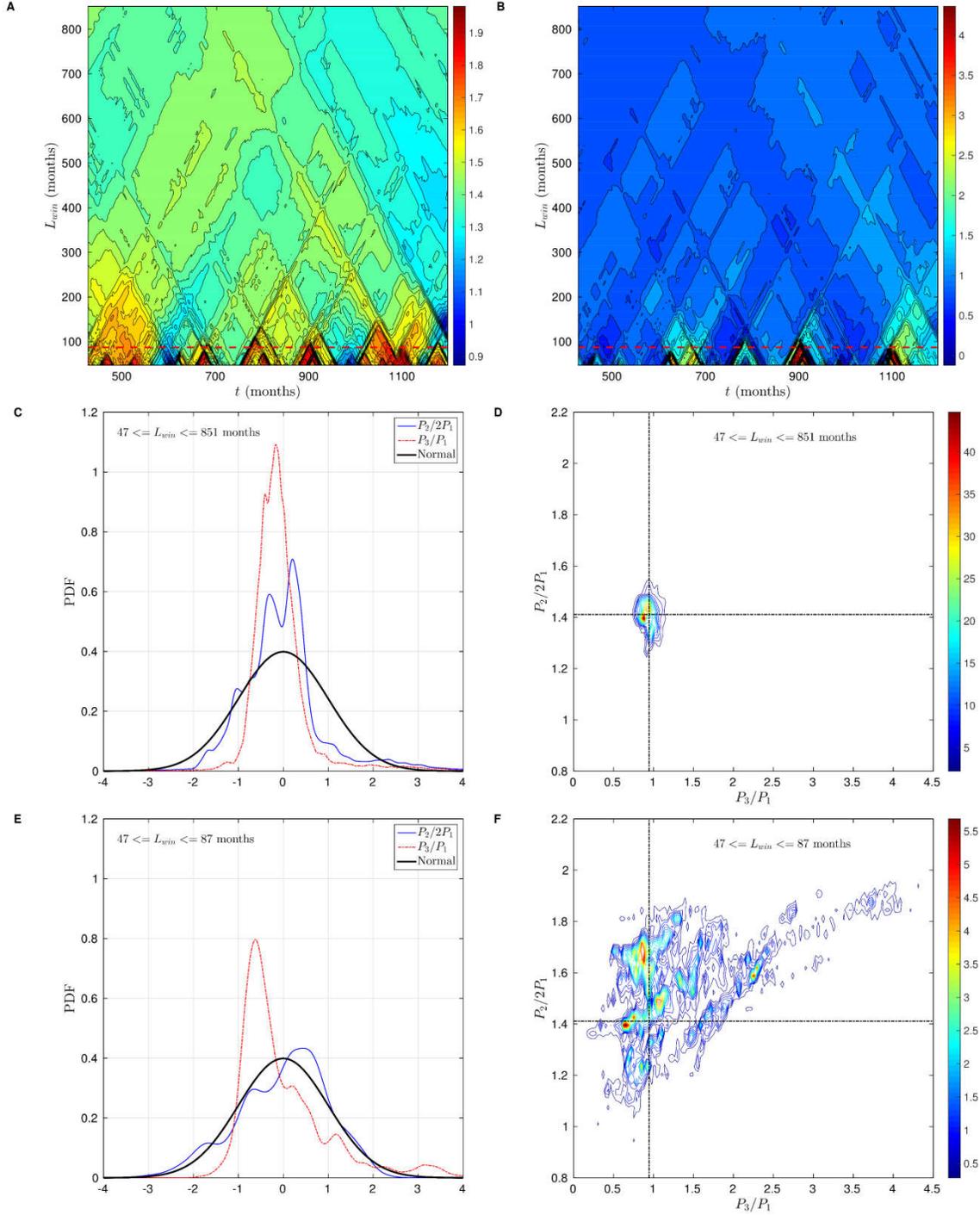

**Fig. S3. Variations of the ratios** $P_2/2P_1$ **and** $P_3/P_1$ **with time.** Contours of **A,** $P_2/2P_1$ and **B,** $P_3/P_1$ plotted on the plane defined by the time coordinate $t$ and the window size $L_{win}$ ranging from 47 to 851 months. Dashed horizontal line in **A** and **B** shows $L_{win}=87$ months. **C,** and **E,** show the normalized PDFs of ratios $P_2/2P_1$ and $P_3/P_1$ for window size ranges $47 \leq L_{win} \leq 851$ and $47 \leq L_{win} \leq 87$ months respectively. Also shown in each plot is the PDF of the normal distribution. **D,** and **F,** show the contours of the joint probability density function for $P_2/2P_1$ and $P_3/P_1$ corresponding to **C** and **E** respectively. Dashed-dotted lines in **D** and **F** show the values $P_2/2P_1 = 1.4113$ and $P_3/P_1 = 0.9433$ obtained from the complete ENSO slow manifold data.

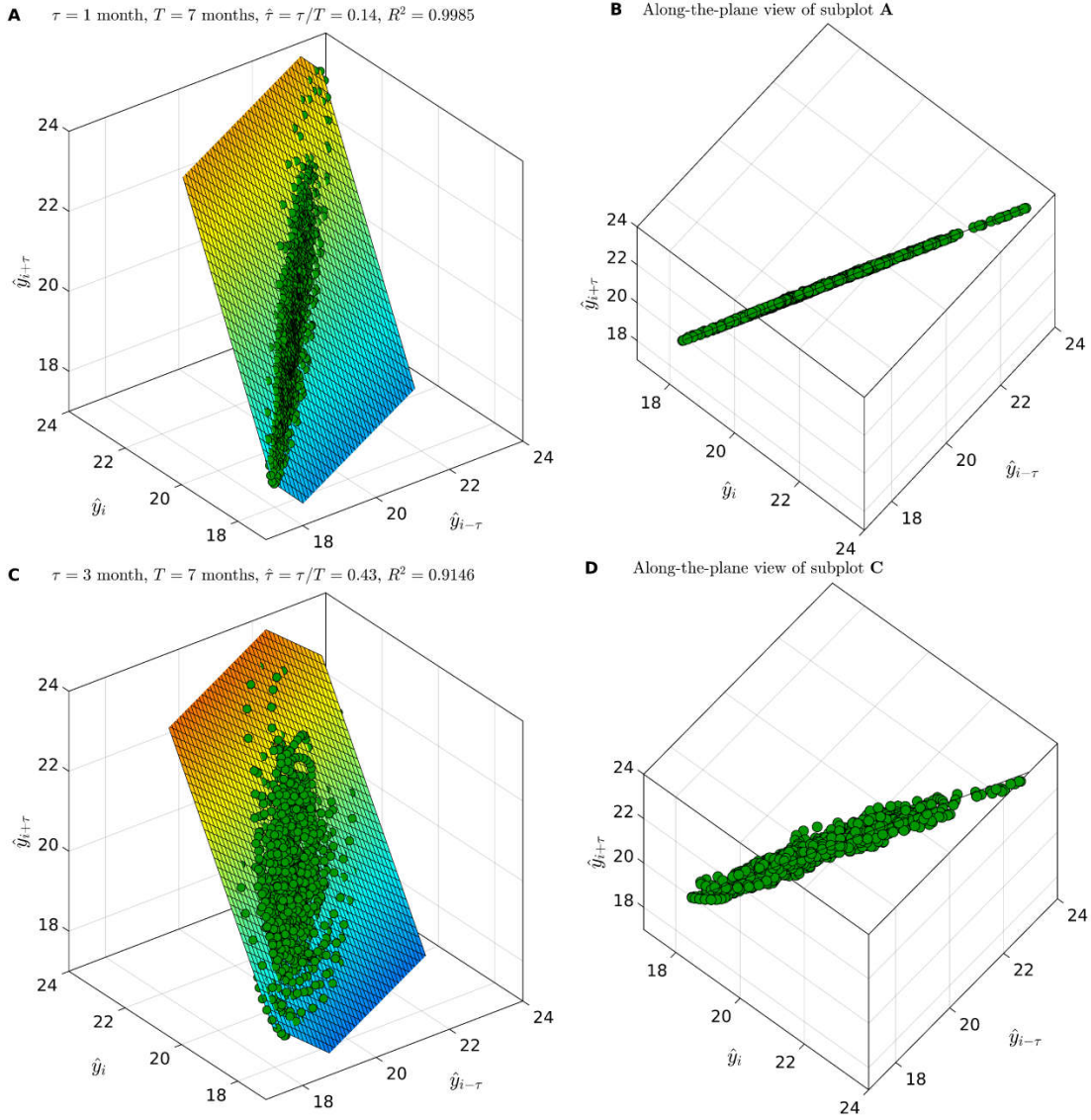

**Fig. S4. Time-invariant, deterministic, linear model of the ENSO slow manifold $y$.** Coordinates of the three-dimensional phase space are $\hat{y}_i$, $\hat{y}_{i-\tau}$ and $\hat{y}_{i+\tau}$ where $\hat{y} = y + 20$. Each point in this space represents the state of the ENSO slow manifold. Trajectories are shown for dimensionless time delay of **A,** $\hat{\tau} = \tau/T = 0.14$ and **C,** $\hat{\tau} = 0.43$. The system evolution in **A** is fairly well-confined to a flat plane (fitted by the least-squares method) given by Eq. (4) as seen in **B**. However as $\hat{\tau}$ is increased to 0.43, it is clear that the trajectory in **C** and **D** does not remain confined to a flat plane. **B** and **D** show along-the-plane views of **A** and **C** respectively and clearly show how nonlinearity takes the trajectory out of the plane for larger delays. This behavior suggests the need of a higher-dimensional phase-space description such as that of Fig. 2 of the main paper.

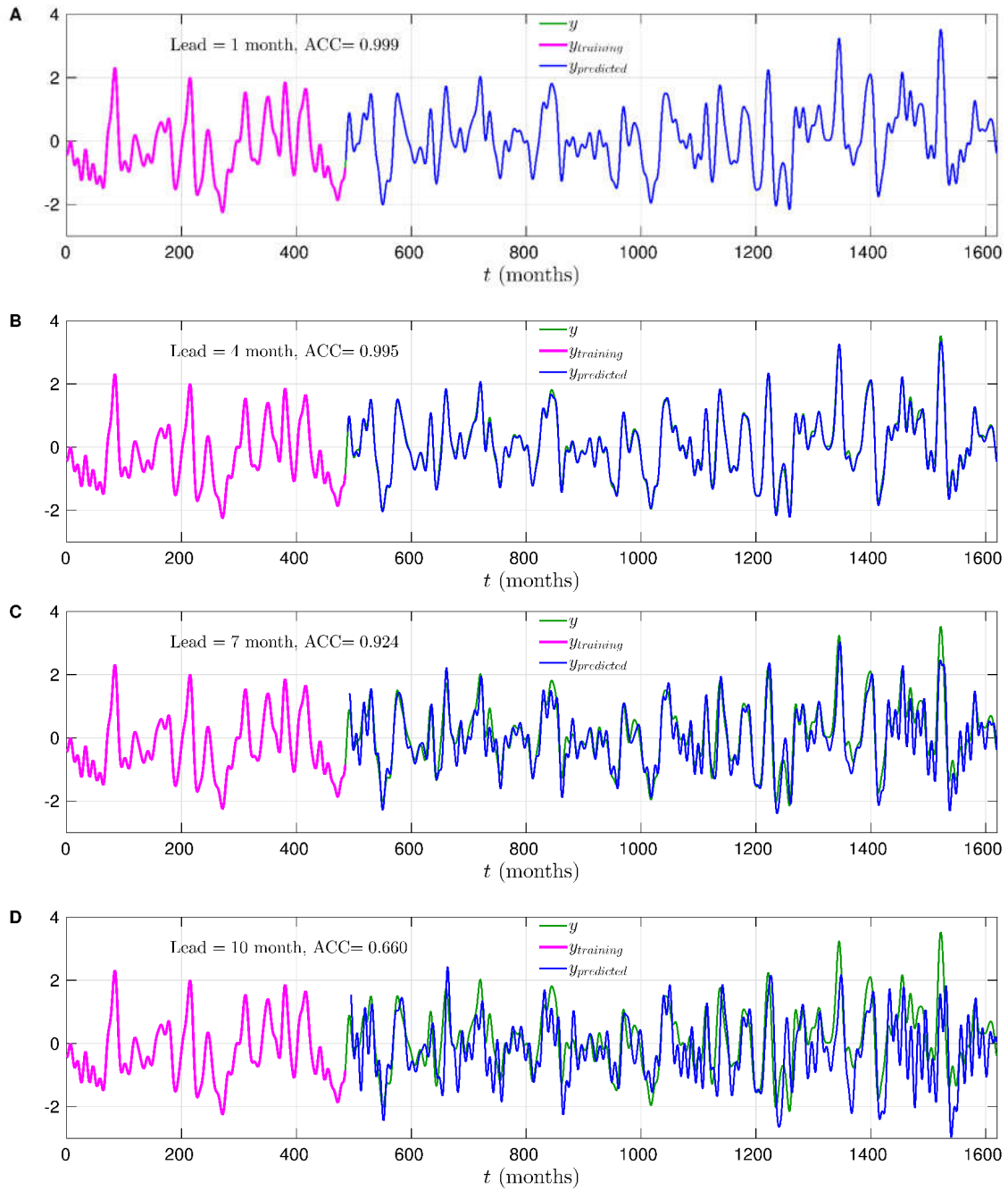

**Fig. S5. Sample nonlinear-model hindcast timeseries of the ENSO slow manifold** $y$. Lead time of **A,** 1 month, **B,** 4 months, **C,** 7 months and **D,** 10 months are shown. Thick magenta line, up to $t = 485$ months (40.5 years) i.e. 30% of the total of 1620 datapoints, shows data used to train the nonlinear model (Eq. 2) for obtaining the required coefficients. DC-shift of 20 is used i.e. $\hat{y} = y + 20$. Green solid line shows the original $y$ timeseries. Blue solid line shows hindcasts generated by Eq. (2).

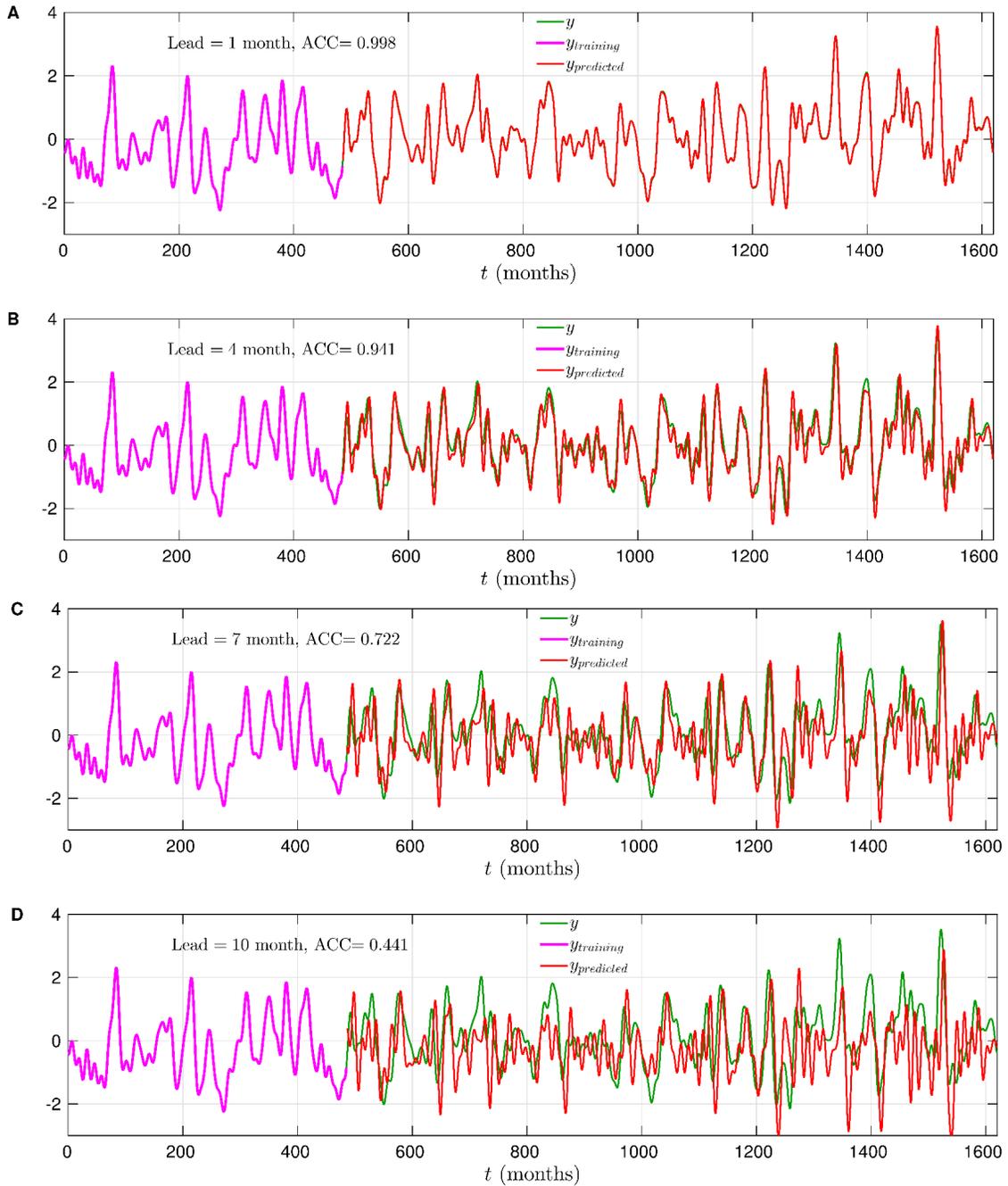

**Fig. S6. Sample linear-model hindcast timeseries of the ENSO slow manifold** $y$. Lead time of **A,** 1 month, **B,** 4 months, **C,** 7 months and **D,** 10 months are shown. Thick magenta line, up to $t = 485$ months (40.5 years) i.e. 30% of the total of 1620 datapoints, shows data used to train the linear model (Eq.4) for obtaining the required coefficients. DC-shift of 20 is used i.e. $\hat{y} = y + 20$. Green solid line shows the original $y$ timeseries. Red solid line shows hindcasts generated by Eq. (4).

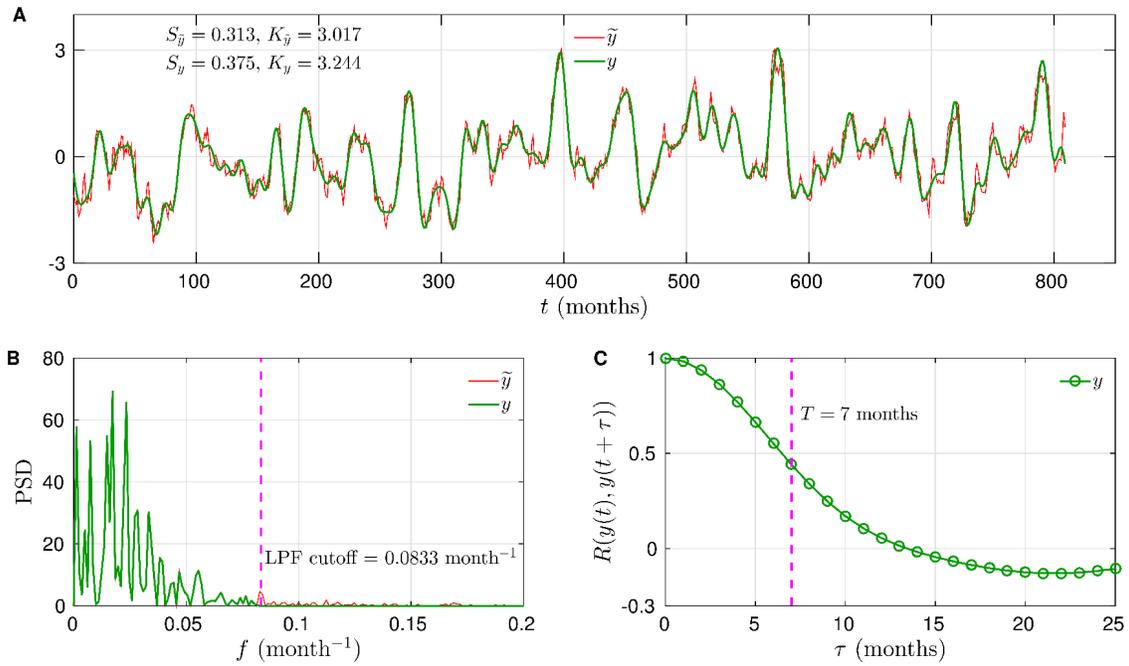

**Fig. S7. Timeseries of the ENSO index MEI and the corresponding ENSO slow manifold. A,** Timeseries of the ENSO index MEI (*43*) denoted as $\widetilde{y}$. The MEI extends from January 1950 ($t=0$) to June 2017 ($t=809$) i.e. more than 67 years with effectively one data point every month i.e. 810 datapoints. ENSO slow manifold is denoted by $y$. For other details see legend of Fig. 1 of the main text.

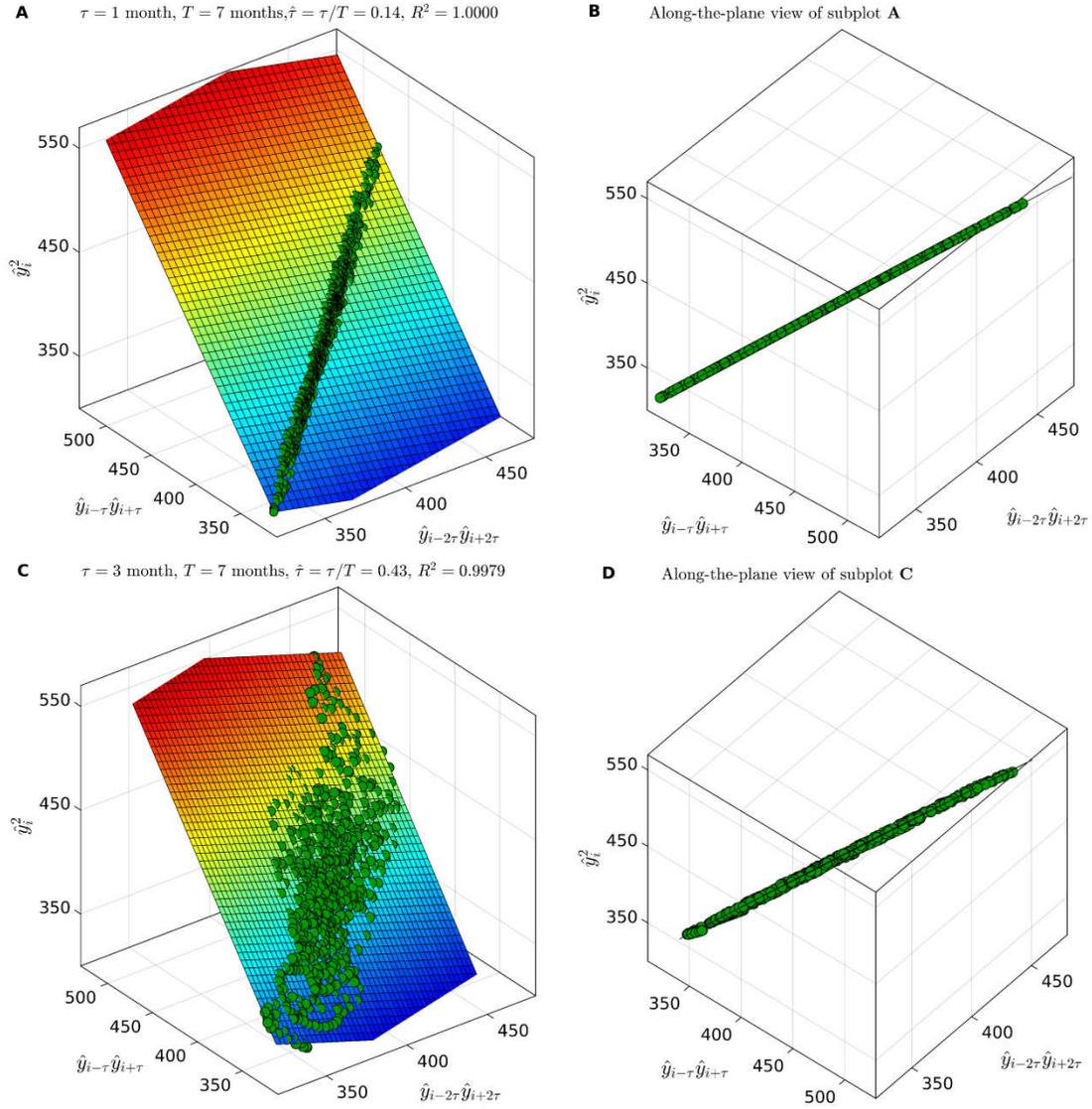

**Fig. S8. Time-invariant, deterministic, nonlinear model of the ENSO slow manifold $y$ constructed from the ENSO index MEI.** For detailed description, see legend of Fig. 2 of the main text. Trajectory of the slow manifold is shown for dimensionless time delays of **A,** $\hat{\tau} = \tau/T = 0.14$ and **C,** $\hat{\tau} = 0.43$.

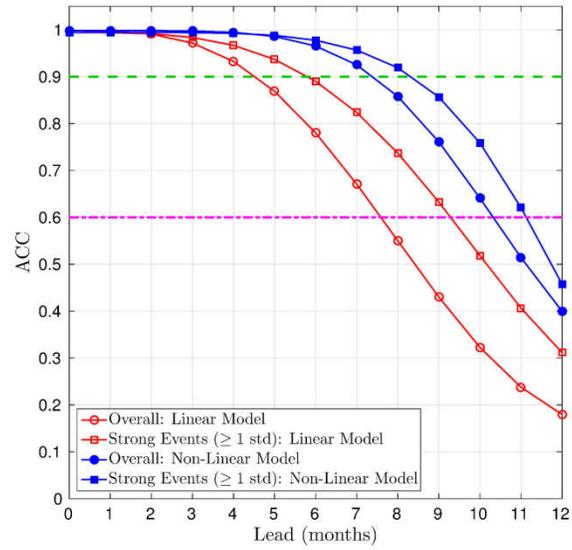

**Fig. S9. Hindcast prediction skill of the present nonlinear and linear models for the ENSO index MEI.** For detailed description, see legend of Fig. 3 of the main text.

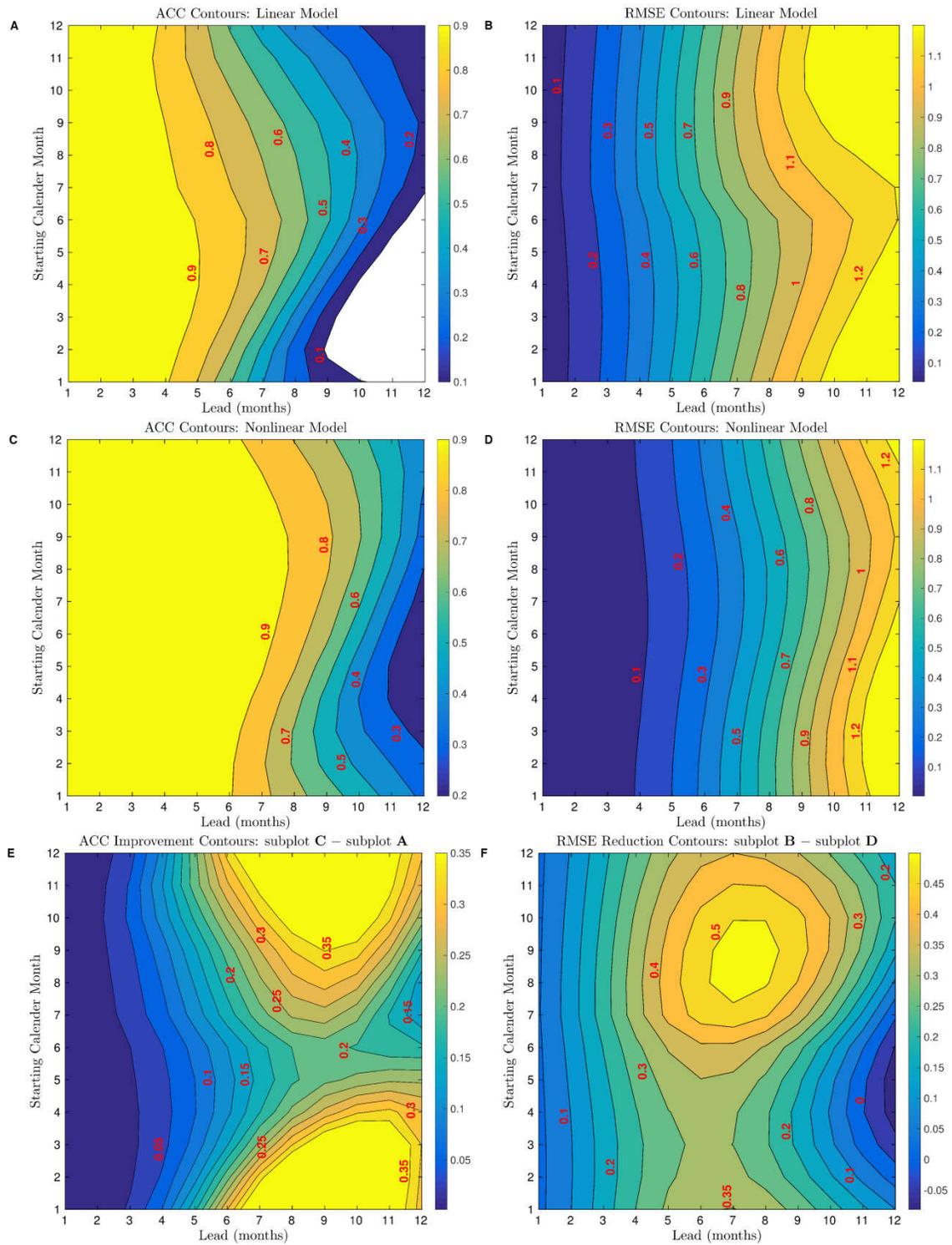

**Fig. S10. Seasonality of the hindcasts generated by the present linear and nonlinear models for ENSO index MEI.** For detailed description, see legend of Fig. 4 of the main text.

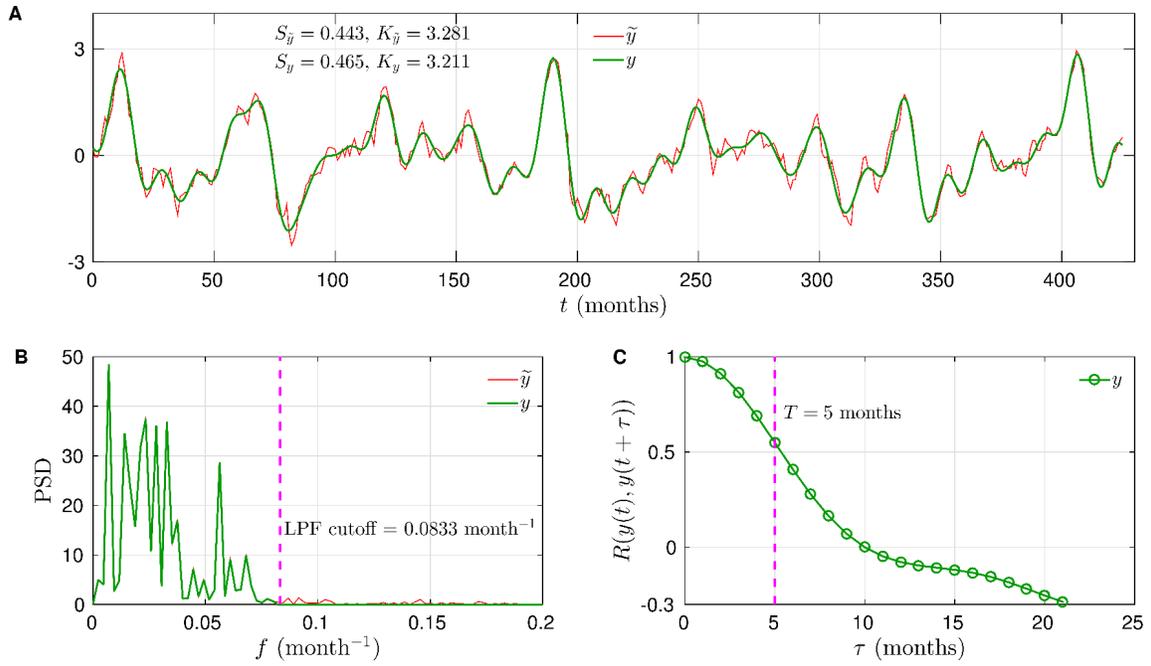

**Fig. S11. Timeseries of the ENSO index Niño3.4 and the corresponding ENSO slow manifold. A,** Timeseries of the ENSO index Niño3.4 denoted as $\widetilde{y}$. Niño3.4 extends from January 1982 ($t=0$) to June 2017 ($t=425$) i.e. more than 35 years with one data point every month i.e. 426 datapoints. ENSO slow manifold is denoted by $y$. For other details see legend of Fig. 1 of the main text.

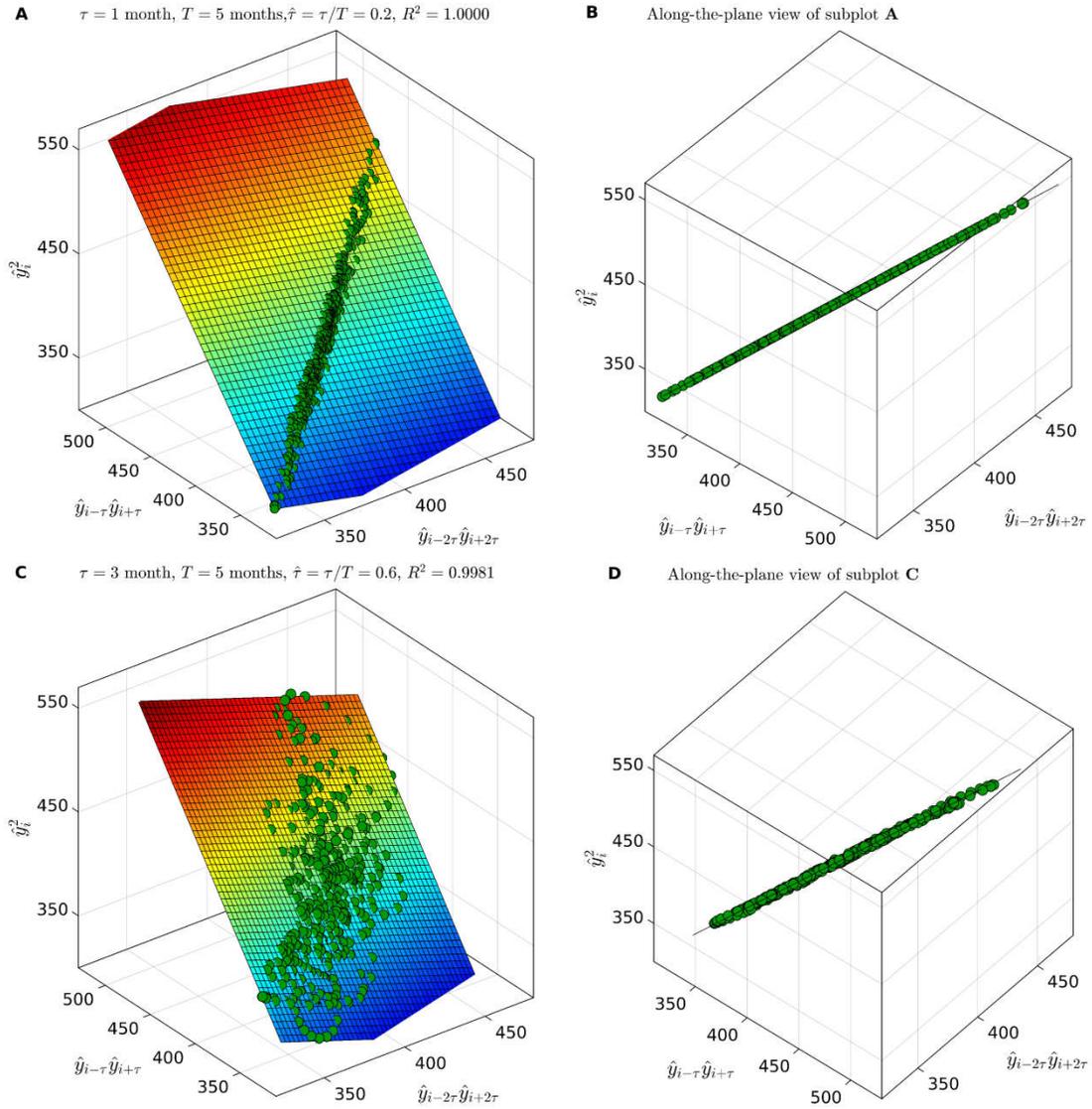

**Fig. S12. Time-invariant, deterministic, nonlinear model of the ENSO slow manifold $y$ constructed from the ENSO index Niño3.4.** For detailed description, see legend of Fig. 2 of the main text. Trajectory of the slow manifold is shown for dimensionless time delays of **A,** $\hat{\tau} = \tau/T = 0.2$ and **C,** $\hat{\tau} = 0.6$.

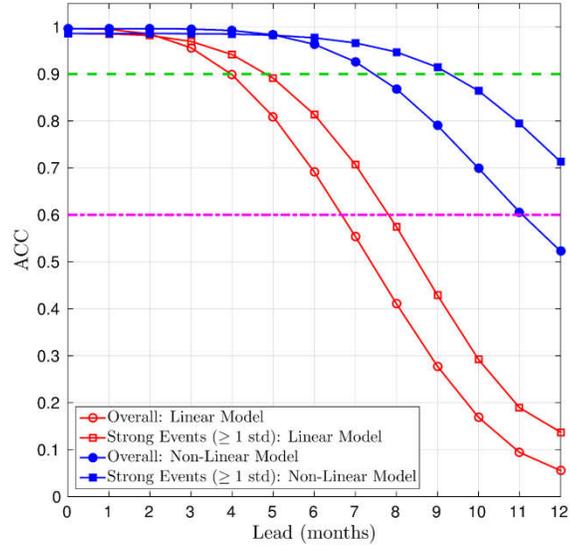

**Fig. S13. Hindcast prediction skill of the present nonlinear and linear models for the ENSO index Niño3.4.** For detailed description, see legend of Fig. 3 of the main text.

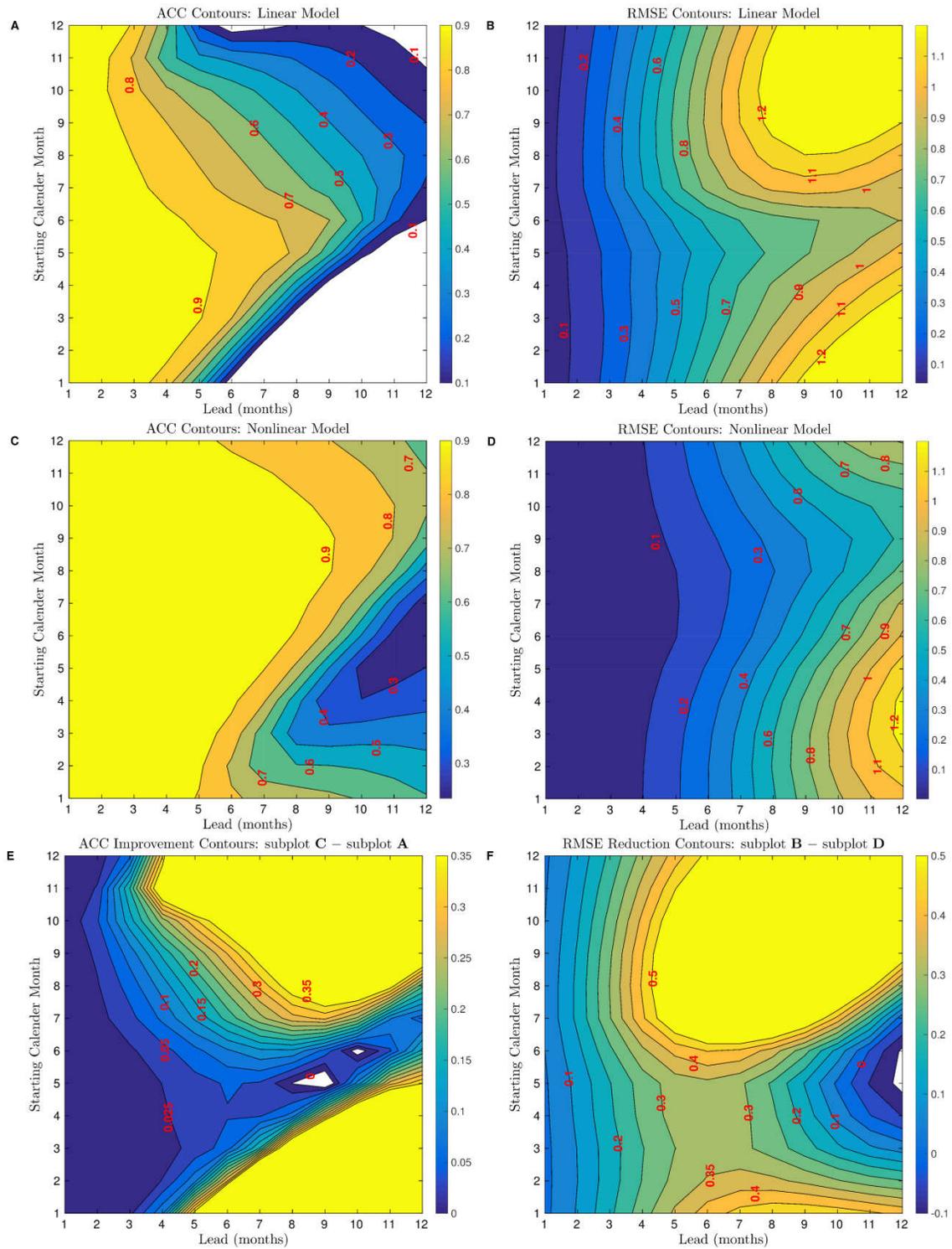

**Fig. S14. Seasonality of the hindcasts generated by the present linear and nonlinear models for the ENSO index Niño3.4.** For detailed description, see legend of Fig. 4 of the main text.

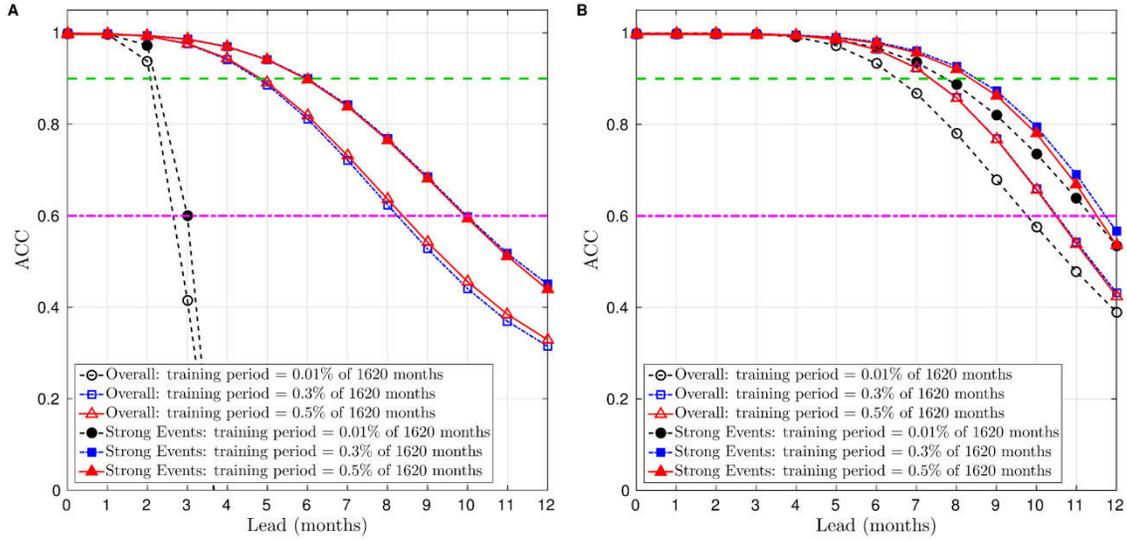

**Fig. S15. Effect of the training period of the linear and nonlinear models on hindcast prediction skill.** Anomaly Correlation Coefficient (ACC) of $y$, between predicted and original datapoints is plotted against lead time for **A,** linear model (Eq. 4) and **B,** nonlinear model (Eq. 2). Curves for different training periods are shown. Strong events indicate slow manifold values exceeding one standard deviation. For reference, horizontal dashed line shows ACC = 0.9 and dashed-dotted line shows ACC = 0.6.